\documentclass[aps,prd,reprint,superscriptaddress,amsmath,amssymb,nofootinbib]{revtex4-1}
\usepackage{xcolor,comment}
\usepackage{multirow}
\usepackage{graphicx}
\usepackage{dcolumn}
\usepackage{bm}
\usepackage{color}
\usepackage{booktabs}
\usepackage{makecell}
\usepackage{hyperref}
\newcommand{\be}{\begin{equation}}
\newcommand{\ee}{\end{equation}}
\newcommand{\bea}{\setlength\arraycolsep{2pt} \begin{eqnarray}}
\newcommand{\eea}{\end{eqnarray}}

\begin{document}

\title{Cosmological implications of two mock TianQin standard siren catalogs generated under different background cosmological models}

\author{Jun-Xian Li}
\email{lijx389@mail2.sysu.edu.cn}
\affiliation{School of Physics and Astronomy, Sun Yat-Sen University, Zhuhai 519082, China.}

\author{Shuang Wang}
\email{wangshuang@mail.sysu.edu.cn}
\thanks{(corresponding author)}
\affiliation{School of Physics and Astronomy, Sun Yat-Sen University, Zhuhai 519082, China.}

\author{Liang-Gui Zhu}
\email{lianggui.zhu@ustc.edu.cn}
\affiliation{Department of Astronomy, University of Science and Technology of China, Hefei, Anhui 230026, China.}
\affiliation{School of Astronomy and Space Sciences, University of Science and Technology of China, Hefei, Anhui 230026, China.}


\begin{abstract}
Space-borne gravitational wave (GW) detectors such as TianQin are expected to detect GW standard sirens for cosmological parameter estimation. 
All previous studies of TianQin adopted the $\Lambda$ cold dark matter ($\Lambda$CDM) model to generate mock standard siren data. 
In this work, we quantitatively investigate how the choice of background cosmological model affects mock TianQin standard siren (TQSS) catalogs and the corresponding cosmological implications. 
Adopting the $\Lambda$CDM model and a redshift-binned dark energy (DE) reconstruction model (the wbin model), we consider the bright standard sirens from massive black hole binary mergers and generate two mock TQSS catalogs, i.e. TQSS1 and TQSS2, respectively. 
We investigate the difference of the catalogs and further explore their cosmological implications within the Chevallier-Polarski-Linder (CPL) model, both individually and in combination with Dark Energy Spectroscopic Instrument Data Release 2 (DESI DR2) baryon acoustic oscillation (BAO) data, $\textit{Planck}$ 2018 cosmic microwave background (CMB) data, and Dark Energy Survey Year 5 (DESY5) type Ia supernova (SN Ia) data. 
We find that there are only small differences between TQSS1 and TQSS2 catalogs, with a maximum relative difference of $1.48\%$ in the distance-redshift relation. 
In addition, either the TQSS1 or TQSS2 catalog alone gives cosmological constraints similar to those obtained from SN data, implying that these two catalogs can serve as a complement to mainstream cosmological observations. 
Most importantly, we find that both TQSS1+DESI DR2+CMB+DESY5 and TQSS2+DESI DR2+CMB+DESY5 are still expected to favor dynamical DE, with these two combinations showing a preference for dynamical DE at $2.3\sigma$ and $3.8\sigma$, respectively. 
Adopting different massive black hole seed models does not impact the cosmological implications of the mock TQSS catalogs. 
Our study indicates the importance of considering different cosmological models in the standard siren forecasts for DE.
\end{abstract}

\maketitle

\section{Introduction}
Gravitational wave (GW) observations are expected to serve as a powerful and independent probe to explore the Universe \cite{Sathyaprakash:2009xs, Bian:2025ifp, Jin:2025dvf, Zhao:2026uvn}.
In particular, GW standard siren is an important method for directly inferring the luminosity distance from the amplitude of the GW waveform \cite{Schutz:1986gp, Holz:2005df}. 
By combining luminosity distance with redshift information, GW standard sirens provide a promising probe to investigate the cosmic expansion history and to constrain cosmological parameters.
For example, observed by the Advanced LIGO--Virgo detectors \cite{LIGOScientific:2017vwq, LIGOScientific:2017ync}, the GW170817 event \cite{LIGOScientific:2017adf} has demonstrated the feasibility of using GW standard sirens for cosmological measurements.
Currently, the LIGO--Virgo--KAGRA (LVK) detector network \cite{KAGRA:2013rdx} continues to accumulate GW events, while the next-generation ground-based GW detectors \cite{VIRGO:2023elp, Punturo:2010zz, Reitze:2019iox} are expected to improve the sensitivity and localization capability of GW observations.

In contrast to ground-based GW detectors, space-borne GW detectors such as the Laser Interferometer Space Antenna (LISA) \cite{LISA:2017pwj}, Taiji \cite{Hu:2017mde}, and TianQin \cite{TianQin:2015yph, TianQin:2020hid} are designed to detect GW signals at much lower frequencies. 
In the frequency range $10^{-4}\,$Hz -- 1Hz, TianQin can detect three classes of GW standard sirens: massive black hole binaries (MBHBs) ($10^5-10^8M_\odot$) \cite{ Wang:2019ryf, Feng:2019wgq, Zhu:2021aat}, stellar-mass binary black holes (SBBHs) ($\sim 10 M_\odot$) \cite{ Liu:2020eko, Liu:2021yoy}, and extreme-mass-ratio inspirals (EMRIs) \cite{Fan:2020zhy, Zi:2021pdp, Fan:2024nnp}.
These TianQin standard sirens (TQSS) can be used to perform cosmological parameter estimation. For example, in Ref.~\cite{Gao:2024uqc}, Gao et al. analyzed the impact of various parameter changes on the estimation of parameters for MBHB systems using TianQin--LISA network.
In Ref.~\cite{Zhu:2021bpp}, Zhu et al. conducted a comprehensive study on the TianQin's ability to constrain the Hubble constant using mock standard sirens from SBBH.
In Ref.~\cite{Zhu:2024qpp}, Zhu et al. demonstrated that localization information can be used to infer the formation channel of EMRIs, thereby improving the precision of cosmological parameters.
Based on the mock data of standard sirens, previous studies have conducted detailed evaluations of the TianQin's detection capability (see \cite{ Li:2024rnk, Hu:2024vvu, Luo:2025ewp, Luo:2025sos} for reviews).

To obtain the mock standard siren data, one must assume a background cosmological model. 
It must be emphasized that all the previous studies of TianQin \cite{Liu:2020eko, Liu:2021yoy, Wang:2019ryf, Feng:2019wgq, Zhu:2021aat, Fan:2020zhy, Zi:2021pdp, Fan:2024nnp, Zhu:2021bpp, Gao:2024uqc, Zhu:2024qpp} adopted the $\Lambda$ cold dark matter ($\Lambda$CDM) model as the background cosmological model \footnote{Many other studies also generated mock standard siren catalogs using the $\Lambda$CDM model as the background cosmological model, see Refs.~\cite{Teixeira:2023zjt, Han:2023exn, Dong:2024bvw, Giare:2024syw, Xiao:2024nmi, Santos:2025wiv, Dong:2025ikq, Liu:2025sgz, Yin:2026gss, Wu:2026mox, Tagliazucchi:2026dpr, Song:2026kii, Dong:2026uxr} for details.}.
However, this choice has three potential issues.
Firstly, if the GW data are simulated in the framework of $\Lambda$CDM, they would naturally favor the $\Lambda$CDM model, which may introduce bias when the data are used to constrain alternative cosmological models. 
Secondly, the $\Lambda$CDM model has encountered a serious challenge known as the Hubble tension \cite{DiValentino:2021izs, Abdalla:2022yfr, Rong-Gen:2023dcz, Efstathiou:2024dvn, CosmoVerseNetwork:2025alb, Cai:2026swf, Li:2026asg}, namely, a discrepancy at the $\sim 5\sigma$ confidence level (CL) between early- and late-time measurements of the Hubble constant $H_0$.
Thirdly, recent studies of the Dark Energy Spectroscopic Instrument (DESI) \cite{DESI:2024mwx, DESI:2025zgx} indicate that its observations show a significant deviation from the $\Lambda$CDM model in the late-time Universe \footnote{There are many discussions on this deviation, see recent Refs. \cite{Cheng:2025yue, Li:2026xaz, Wang:2026rkx, Feng:2026pzs, Jiang:2026nug, Paliathanasis:2026vhi, Gomez-Valent:2026ept, Li:2026hwq, Dinda:2026ktu, Toda:2026yum, Wu:2026qog, Du:2026qtq, Li:2026ldf, Zhou:2026dti, Huang:2026szy, SolaPeracaula:2026lyk, Huang:2026khc, Yang:2026vti, Basilakos:2026atu, Ishak:2026wdu, Rippentrop:2026kpb, Leizerovich:2026qie, Wang:2026hjy, Argudo-Panes:2026qkw, Zhang:2026nap, Huang:2026igp} for more details.}.
Therefore, it is important to take into account different background cosmological models in the simulation of standard siren data.

Instead of adopting a specific model, many efforts have been devoted to the model-independent reconstruction of dark energy (DE) \cite{Huterer:2002hy, Shafieloo:2005nd, Holsclaw:2010sk, Seikel:2012uu, Zhao:2012aw, Nesseris:2012tt, Shafieloo:2012yh, Huang:2009rf, Wang:2010vj, Luo:2018yvq}.
Based on these DE reconstruction methods, numerous studies \cite{Luongo:2024fww, Mukherjee:2024ryz, DESI:2024aqx, Orchard:2024bve, Jiang:2024xnu, Reboucas:2024smm, Pang:2024qyh, Liu:2024yib, Yang:2025kgc, Bansal:2025ipo, Gao:2025ozb, Ormondroyd:2025exu, DESI:2025wyn, Wang:2025vfb, Li:2025ula, Cheng:2025cmb, Gonzalez-Fuentes:2025lei, Wang:2026kbg, Wang:2026sqy, Kessler:2026dbi, Borghetto:2026ytv} have shown that the DESI data favor a dynamical evolving DE.
For example, in Ref.~\cite{Li:2025ops}, we employed the redshift binning method \cite{Huterer:2002hy} to analyze the DESI data and found that DE should evolve with redshift, with the equation of state (EoS) deviating from $w=-1$ at least at the $2.13\sigma$ CL.
Compared with the $\Lambda$CDM model, the redshift binning method has two main advantages: 
(1) it does not assume a specific functional form of the DE EoS; 
(2) it can reveal the details of DE's dynamical evolution.
Therefore, in this paper, we adopt the redshift binning method to reconstruct the DE EoS, and we refer to this background cosmological model as the wbin model hereafter.

In this work, we mainly focus on investigating two mock TianQin standard siren catalogs based on different background cosmological models.
In particular, we consider the $\Lambda$CDM model and the wbin model as the background cosmological models, respectively.
To generate the mock standard siren data, we first constrain these two background cosmological models using a data combination of DESI DR2 baryon acoustic oscillation (BAO), $\textit{Planck}$ 2018 cosmic microwave background (CMB), and DESY5 type Ia supernova (SN Ia).
It should be noted that, for the CMB data, we use the full CMB likelihoods rather than the Planck 2018 distance priors used in our previous work \cite{Li:2025ops}.
Compared with the distance priors approach, the use of full CMB likelihoods can provide tighter parameter constraints and help break the parameter degeneracies.
After constraining the background cosmological models, we then adopt the corresponding cosmological parameters as input to generate the mock data, where TQSS1 is based on the $\Lambda$CDM model and TQSS2 is based on the wbin model.
Combining the TQSS1 and TQSS2 catalogs with the DESI DR2 BAO, $\textit{Planck}$ 2018 CMB, and DESY5 SN observational data, we obtain the cosmological constraints in the theoretical framework of the Chevallier-Polarski-Linder (CPL) model \cite{Chevallier:2000qy, Linder:2002et}, and further investigate the corresponding cosmological implications. 

The main novelty of this paper is adopting the wbin model as the background cosmological model to generate the mock TianQin standard siren catalogs. 
It should be emphasized that this treatment has not been adopted in previous studies of TianQin.

The structure of this paper is as follows. 
In Section~\ref{sec:background-setup}, we introduce the theoretical foundation for the cosmological models and present the corresponding cosmological constraints.
In Section~\ref{sec:mockGW}, we describe the simulation method used to generate TianQin GW standard siren catalogs. 
In Section~\ref{sec:results}, we present the cosmological implications of the TQSS catalogs, and discuss the corresponding results.
In Section~\ref{sec:Summary}, we provide a brief summary.
Throughout this paper, the spatially flat Friedmann--Lemaître--Robertson--Walker (FLRW) universe is adopted and the subscript '0', denoted as present day, is dropped on the density parameters.

\section{Theoretical foundation and cosmological constraint}
\label{sec:background-setup}
In this section, we first introduce the theoretical foundation of cosmological models.
We then briefly outline the mainstream observational data sets and statistical methodologies used in this paper.
Finally, we present the corresponding cosmological constraint results for the background cosmological models. 

\subsection{Cosmological models}
\label{subsec:mi_recon}
In GW cosmology, the luminosity distance is a key quantity, which is given by
\begin{equation}
\label{eq:lum_dist}
    d_L = \frac{c(1+z)}{H_0} \int^z_0 \frac{dz'}{H(z)/H_0},
\end{equation}
where $c$ is the speed of light, $z$ is the redshift, $H(z)$ is the Hubble parameter. 
Governed by the Friedmann equation, the $H(z)$ can be expressed as
\begin{equation}
\label{eq:Friedmnn}
    \frac{H(z)}{H_0} = \sqrt{ \Omega_r (1+z)^4 + \Omega_m (1+z)^3  + \Omega_{\rm DE} f(z) },
\end{equation}
where $\Omega_r$, $\Omega_m$, and $\Omega_{\rm DE}$ refer to the present fractional energy densities of radiation, matter, and dark energy, respectively. 
In a flat universe, $\Omega_{\rm DE}=1-\Omega_r -\Omega_m$. 
$f(z)$ is the effective (normalized) DE density, which is related to the equation of state (EoS) parameter $w(z)$ as
\begin{equation}
\label{eq.fz}
    f(z)\equiv \frac{\rho_{de}(z)}{\rho_{de}(0)}=\text{exp} \left[ 3\int_0^z \frac{1+w(z')}{1+z'} dz' \right].
\end{equation}

For the first background cosmological model, we consider the $\Lambda$CDM model. The EoS is $w=-1$; therefore the corresponding DE density is a constant, giving $f(z)=1$. 

For the second background cosmological model, we consider the wbin model, where we adopt the redshift binning method to reconstruct DE.
The key idea of the method is to divide the redshift range into several bins, and to assume the DE EoS $w(z)$ to be piecewise constant within each bin.
Explicitly, the EoS takes the form
\begin{equation}
    w(z)=
    \begin{cases}
         w_1, & 0 < z \leq z_1 \\ 
         w_i, & z_{i-1} < z \leq z_i \quad (i>1)
    \end{cases} \quad .
\end{equation}
Based on the continuity condition, the corresponding $f(z)$ is given by Ref.~\cite{Sullivan:2007pd}, which can be written as
\begin{equation}
    f(z_{n-1} < z \leq z_n) = (1+z)^{3(1+w_n)} \prod_{i=0}^{n-1} (1+z_i)^{3(w_i-w_{i+1})}.
\end{equation}

In Ref.~\cite{Li:2025ops}, we showed that the number of bins does not significantly affect the final reconstruction results.
Moreover, the higher redshift bin is weakly constrained due to the limited data, as the bin increases.
Therefore, we adopt four redshift bins in this work.
The complete external redshift information for GW events beyond $z=3$ is expected to be difficult to obtain \cite{Zhu:2021bpp}.
We thus restrict the reconstruction to $0<z \leq 3$, where the redshift-binned points are $z_i=0.75/1.5/2.25/3.0 \quad (i=1, \cdots, 4)$.

As mentioned above, we adopt the CPL model to discuss the dynamical behavior of DE in Section~\ref{sec:results}, we thus introduce this model here.
The EoS of the CPL model takes the form 
\begin{equation}
\label{eq:cpl_w}
    w(z)=w_0 + w_a \frac{z}{1+z},
\end{equation}
where $w_0$ and $w_a$ are constants. 
The corresponding normalized density is given by
\begin{equation}
\label{eq:cpl_f}
    f(z) = (1+z)^{3(1+w_0+w_a)} \text{exp}\left( -\frac{3w_a z}{1+z} \right).
\end{equation}
When $w_0=-1$, $w_a=0$, the CPL model returns to the $\Lambda$CDM case. 
As for the deceleration parameter $q(z)$, it determines the cosmological redshift at which the universe began to accelerate. 
The $q(z)$ is giving by
\begin{equation}
\label{eq:cpl_q}
    q(z)\equiv -\frac{\ddot{a} a}{\dot{a}^2}=\frac{d \ln H}{d \ln (1+z)} -1,
\end{equation}
where $a=1/(1+z)$ is the scale factor.

\subsection{Mainstream observational data and statistical methods}
\label{subsec:dataset}
Since the GW data simulation requires cosmological parameter values as input, we need to constrain the $\Lambda$CDM model and the wbin model using a combination of mainstream cosmological data sets. 
In the following, we briefly describe the cosmological data sets and statistical methodologies below.

\begin{itemize}
    \item $\textit{Baryon Acoustic Oscillations}$ $\textbf{(BAO)}$. 
    We utilize the DESI DR2 BAO measurements \cite{DESI:2025zgx}, which includes observations of galaxies and quasars \cite{DESI:2025qqy}, as well as Lyman-$\alpha$ tracers \cite{DESI:2025zpo}. 
    These measurements correspond to several key distances: the transverse comoving distance $D_M$, the Hubble distance $D_H$, and the angle-averaged distance $D_V$.
    These BAO distances, reported in Table IV of Ref.~\cite{DESI:2025zgx}, are expressed in units of the comoving sound horizon at the drag epoch, $r_d$.
    Assuming the standard pre-recombination physics, we adopt the approximation formula of $r_d$ used by DESI \cite{DESI:2025zpo}.
    The treatment of the covariance matrices and correlation coefficients of the DESI BAO data can be found in Ref.~\cite{Li:2024bwr}.
    The likelihood is publicly available in the Bayesian analysis code Cobaya.
    We denote this dataset as $\texttt{DESI DR2}$.

    \item $\textit{Cosmic Microwave Background}$ $\textbf{(CMB)}$. 
    We make use of the CMB temperature (TT), polarization (EE), and cross (TE) power spectra from the $\textit{Planck}$ 2018 data release \cite{Planck:2018vyg}.
    Specifically, we adopt the $\texttt{Commander}$ and $\texttt{simall}$ likelihoods \cite{Planck:2019nip} for low-$\ell$ TT and EE.
    For high-$\ell$ TTTEEE, we adopt the $\texttt{CamSpec}$ PR4 likelihood \cite{Efstathiou:2019mdh, Rosenberg:2022sdy} .
    In addition, we also incorporate the $\textit{Planck}$ PR4 NPIPE lensing likelihood \cite{Carron:2022eyg} in all our analyses. 
    We denote this dataset as $\texttt{CMB}$.

    \item $\textit{Type Ia Supernovae}$ $\textbf{(SNe Ia)}$.
    We use the Dark Energy Survey Year 5 (DESY5) compilation of SNe Ia measurements. 
    The DESY5 compilation \cite{DES:2024jxu} contains 1635 photometrically classified SNe Ia covering the range $0.1<z<1.3$ by the DES Supernova Program during five years, together with 194 low-redshift SNe Ia in $0.025<z<0.1$ that overlap with the PantheonPlus sample \cite{Brout:2022vxf}.
    The likelihood is publicly available in the Bayesian analysis code Cobaya. 
    The covariance matrix, which includes both statistical and systematic errors, is also publicly available.
    We denote this dataset as $\texttt{DESY5}$.
\end{itemize}

For the statistical analyses, we sample the posterior distributions of the cosmological parameters using the Bayesian analysis code $\texttt{Cobaya}$\footnote{\url{https://github.com/CobayaSampler/cobaya}} \cite{Torrado:2020dgo} with its Markov Chain Monte Carlo (MCMC) sampler \cite{Lewis:2013hha, Lewis:2002ah}. 
Theoretical predictions for the relevant cosmological observables are obtained through a modified version of the Boltzmann solver $\texttt{CAMB}$ \cite{Lewis:1999bs}.
Convergence of the MCMC chains is assessed using the Gelman-Rubin criterion \cite{Gelman:1992zz}, requiring $|R-1| < 0.05$. 
Statistical results are obtained by analyzing the MCMC chains with $\texttt{GetDist}$ \cite{Lewis:2019xzd}.
The corresponding one-dimensional and two-dimensional plots are also generated using $\texttt{GetDist}$. 

To measure the statistical significance of the preference for dynamical DE, we make use of the $\texttt{iminuit}$ \cite{dembinski_2025_17565861} minimizer to obtain the $\Delta \chi^2_{\rm MAP}$, where the calculation of the best-fit points is started with the maximum a posteriori (MAP) points.
The $\Delta \chi^2$ is defined as the difference between the best-fit values of the $\Lambda$CDM and CPL models for a given data combination.
According to Wilks' theorem \cite{Wilks:1938dza}, the $\Delta \chi^2_{\rm MAP}$ should follow a $\chi^2$ distribution with two degrees of freedom (dof) under the assumption the null hypothesis ($\Lambda$CDM model) holds, and assuming that errors are Gaussian and correctly estimated.
Therefore, using the cumulative distribution function (CDF)
\begin{equation}
    \mathrm{CDF}_{\chi^2}(\Delta \chi^2_{\rm MAP}|\rm 2\,dof)= \frac{1}{\sqrt{2\pi}} \int^N_{-N} e^{-t^2/2}dt,
\end{equation}
we can translate the $\Delta \chi^2_{\rm MAP}$ into the corresponding frequentist significance $N\sigma$.

In Table \ref{tab:flatprior}, we list the flat priors adopted for the free parameters. 
The sampled parameters are the physical baryon energy density $\Omega_b h^2$, the physical cold dark matter energy density $\Omega_c h^2$, the amplitude of the primordial scalar spectrum $A_s$, the spectral index of the primordial scalar spectrum $n_s$, the optical depth to reionization $\tau$, the angular size of the sound horizon at last scattering $\theta_{\text{MC}}$, and the DE EoS parameter $w_i$.
For the wbin and the CPL models, the flat prior range $[-3,2]$ is adopted for all the DE EoS parameters $w_i$.

\begin{table}
\centering
\renewcommand{\arraystretch}{1.5}
\begin{tabular}{l @{\hspace{2cm}} c}
\hline
\textbf{Parameter} & \textbf{Prior} \\
\hline \hline
$\Omega_\mathrm{b} h^2$ & $[0.005, 0.1]$ \\
$\Omega_\mathrm{c} h^2$ & $[0.01, 0.99]$ \\
$100\theta_\mathrm{MC}$ & $[0.5, 10]$ \\
$\tau$ & $[0.01, 0.8]$ \\
$n_\mathrm{s}$ & $[0.8, 1.2]$ \\
$\log(10^{10} A_\mathrm{s})$ & $[1.61, 3.91]$ \\
$w_i$ & $[-3, 2]$ \\
\hline
\end{tabular}
\caption{Flat priors on the cosmological parameters sampled in this work.}
\label{tab:flatprior}
\end{table}

\subsection{Cosmological constraints on the background cosmological models}
\begin{figure*}
\small
    \centering
        \includegraphics[width=0.48\textwidth]{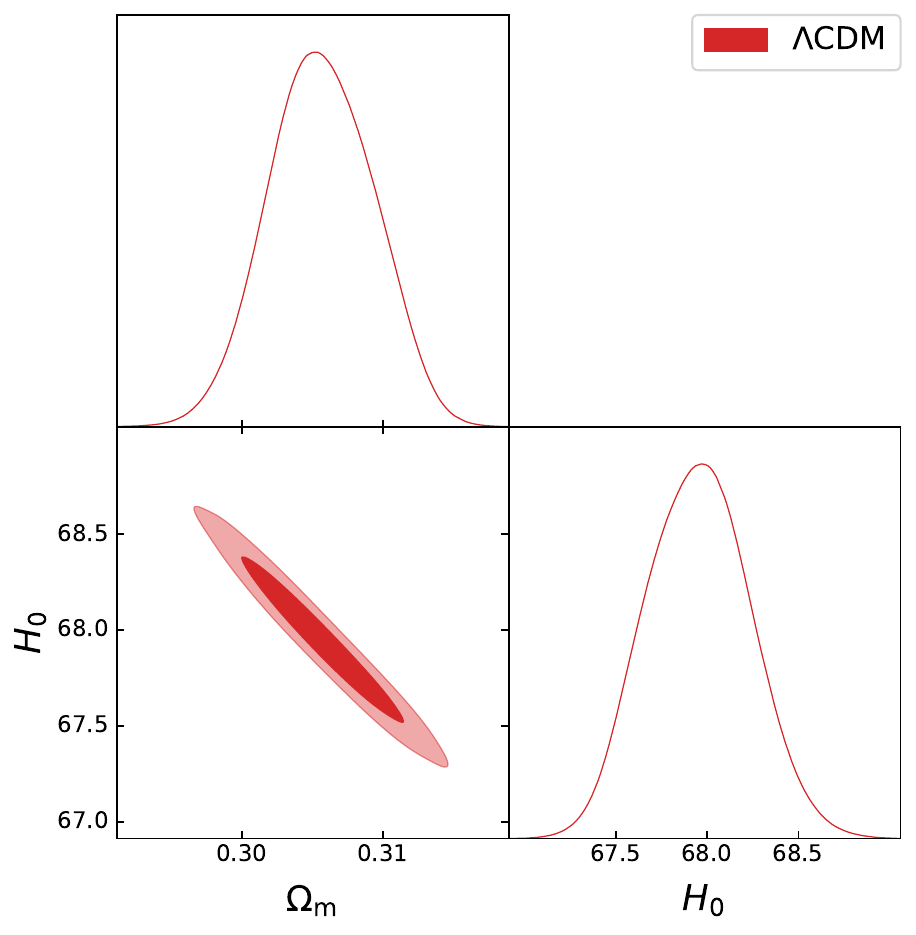}
        \includegraphics[width=0.48\textwidth]{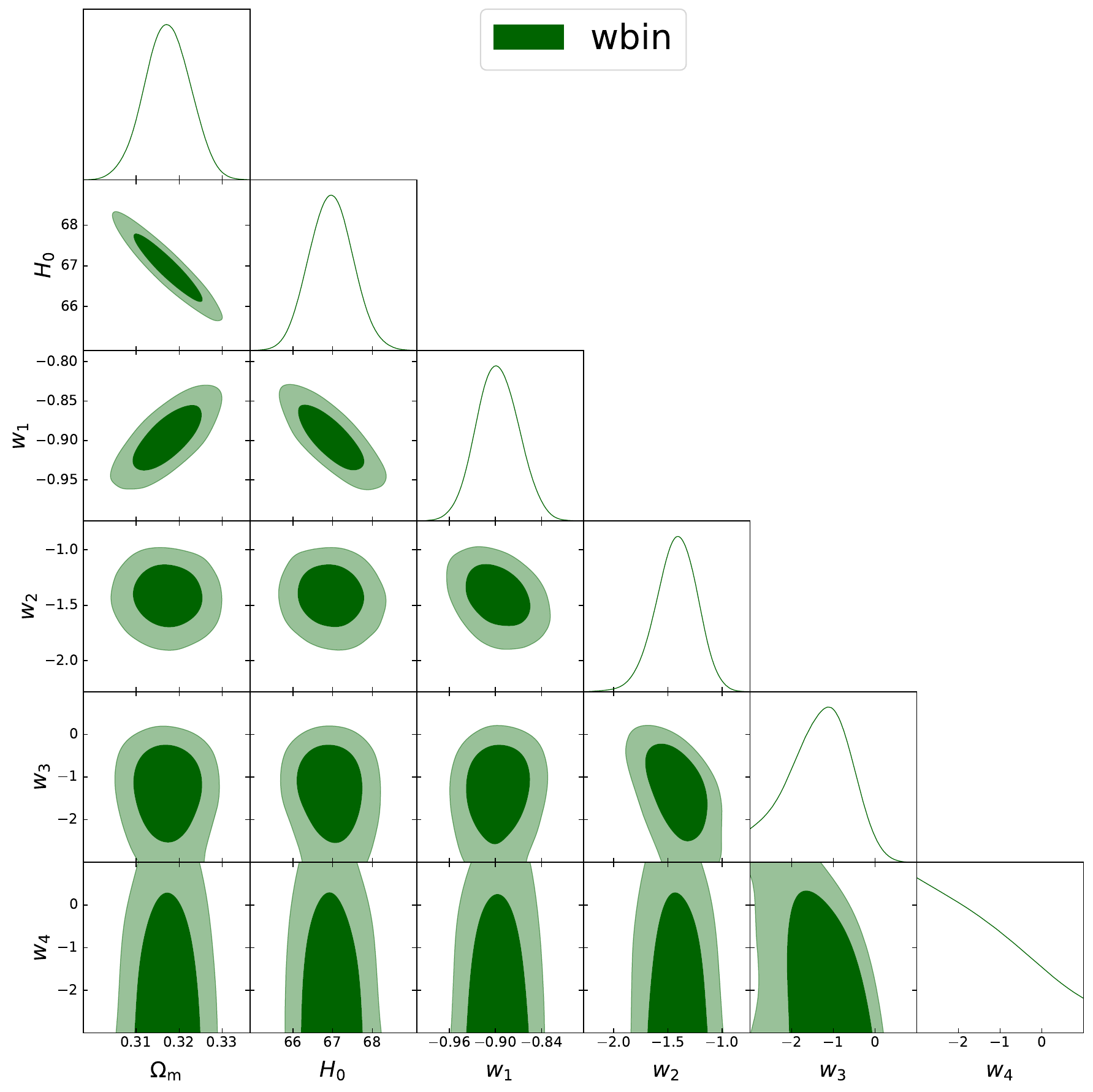}
    \caption{One-dimensional posterior distributions and two-dimensional marginalized contours at 1$\sigma$ and 2$\sigma$ confidence regions for the $\Lambda$CDM model and the wbin model.}
    \label{fig:contour_lcdm&wbin}
\end{figure*}

\begin{table*}
    \centering
    \resizebox{\textwidth}{!}{
    \renewcommand{\arraystretch}{1.5}
    \setlength{\tabcolsep}{4pt}
    \begin{tabular}{lcccccc}
    \hline \hline
    \textbf{Dataset/Model} & $\Omega_m$ & $H_0$[$\mathrm{km\,s^{-1}\,Mpc^{-1}}$] & $w_1$ & $w_2$ & $w_3$ & $w_4$  \\
    \hline
      $\textbf{DESI DR2+CMB+DESY5}$  \\
      $\Lambda$CDM  &  $0.3056 \pm 0.0037$ & $67.95 \pm 0.28$ & $\text{---}$ & $\text{---}$ & $\text{---}$ &$\text{---}$ \\
      wbin  & $0.3172 \pm 0.0052$ & $66.96 \pm 0.55$ & $-0.897 \pm 0.028$ & $-1.421 \pm 0.190$ & $-1.344 \pm 0.719$ & $-1.426 \pm 1.059$    \\
    \hline \hline
    \end{tabular}
}
\caption{Constraints on cosmological parameters, derived from DESI DR2+CMB+DESY5 data combination.}
\label{tab:binningw}
\end{table*}

We constrain the $\Lambda$CDM model and the wbin model using a data combination of DESI DR2+CMB+DESY5.
The one-dimensional posterior distributions and two-dimensional marginalized contours at $1\sigma$ and $2\sigma$ confidence regions are shown in Fig.~\ref{fig:contour_lcdm&wbin}.
The mean values with $1\sigma$ uncertainties are summarized in Table \ref{tab:binningw}.

Since the GW data simulation only requires background cosmological parameters as input, we mainly present the relevant quantities here.

For the $\Lambda$CDM model, the posterior distributions are shown in red on the left panel of Fig.~\ref{fig:contour_lcdm&wbin}.
The mean values are $\Omega_m = 0.3056 \pm 0.0037$ and $H_0 =67.95 \pm 0.28$ $\mathrm{km\,s^{-1}\,Mpc^{-1}}$. 
The EoS of the $\Lambda$CDM model is a constant, $w=-1$.

For the wbin model, the posterior distributions are shown in green on the right panel of Fig.~\ref{fig:contour_lcdm&wbin}.
The matter density parameters and Hubble constant are constrained to $\Omega_m = 0.3172 \pm 0.0052$ and $H_0 =66.96 \pm 0.55$ $\mathrm{km\,s^{-1}\,Mpc^{-1}}$, respectively.
As for the DE parameters, the first bin parameter is tightly constrained to $w_1=-0.897 \pm 0.028$, deviating from the cosmological constant value $w=-1$ at $3.7\sigma$.
In contrast, the higher-redshift bins yield progressively weaker constraints: $w_2= -1.421 \pm 0.190$, $w_3= -1.344 \pm 0.719$, and $w_4= -1.426 \pm 1.059$.
The DE EoS parameter exhibits $w>-1$ in the lowest redshift bin and $w<-1$ in the higher redshift bins, suggesting a possible quintom-like behavior \cite{Cai:2025mas, Ren:2026jyw}.

It should be pointed out that, in our previous work \cite{Li:2025ops}, we constrained the wbin model using the DESI DR2+CMB+DESY5 data combination, where the CMB data were incorporated via the Planck 2018 distance priors.
However, the distance priors are a compressed CMB dataset and are derived under an assumed cosmological model.
Compared with the distance priors approach, the full CMB likelihoods retains more information, which can provide tighter parameter constraints and help break the parameter degeneracies.
Therefore, in this work, we adopt the full CMB likelihood in the DESI DR2+CMB+DESY5 data combination to constrain the background cosmological models considered here.

In the following section, we adopt the mean values in Table \ref{tab:binningw} as input to generate the mock TianQin standard siren catalogs.

\section{Simulation of TianQin GW standard sirens}
\label{sec:mockGW}
In this Section, we first introduce the GW detector response of TianQin and GW signal detection, and then we describe the GW standard siren considered in this paper. 
Thirdly, we describe the GW errors and give the corresponding $\chi^2$ statistics function.

\subsection{TianQin GW detector and signal detection}
For a GW event, the signal measured by a detector is a scalar quantity $h(t)=D^{ij} h_{ij}(t)$ \cite{maggiore2008gravitational}, where $D^{ij}$ is a constant tensor which depends on the detector, $h_{ij}(t)$ is a tensor which describes the GW. 
In the transverse-traceless gauge, $h_{ij}$ has only two independent components $h_+$ and $h_\times$, $h_{xx}=-h_{yy}=h_+$, $h_{xy}=h_{yx}=h_\times$, all other components being zero. 
The GW signal $h(t)$, also called strain amplitude, contains only a certain linear combination of the two components, which is given by \cite{maggiore2008gravitational}:
\begin{equation}
    h(t) = F_+ (\theta,\phi,\psi)h_+(t) + F_\times (\theta,\phi,\psi)h_\times (t),
\end{equation}
where $F_+$ and $F_\times$ are the detector pattern functions, given through contraction of detector tensor and polarization tensor as $F_A=D^{ij}e^A_{ij}$ (with $A=+,\times$ labeling the polarizations). 
$\psi$ is the polarization angle, and $(\theta,\phi)$ are angles describing the location of the source on the sky, relative to the detector.

When a GW signal enters a detector, the performance of detector is characterized by the one-sided power spectral density (PSD) $S_n(f)$.
We take the noise PSD of the TianQin as \cite{TianQin:2015yph, Zhu:2021aat}:
\begin{equation}
    \begin{split}
        S_n(f) = & \frac{1}{L^2}\left[ \frac{4S_a}{(2\pi f)^4} \left( 1+ \frac{10^{-4}\text{Hz}}{f} \right) +S_x \right] \\
        &\times \left[ 1+0.6\left( \frac{2\pi fL}{c}\right)^2 \right],
    \end{split}
\end{equation}
where $L=\sqrt{3} \times 10^8$m is the arm length, $S_a^{1/2}=1\times 10^{-15}$m s$^{-2}$ Hz$^{-1/2}$ is the acceleration noise, and $S_x^{1/2}=1\times 10^{-12}$m Hz$^{-1/2}$ is the positional noise.
In order to deal with the estimation of parameters in the waveform, we employ the Fisher information matrix approach \cite{Finn:1992wt, Cutler:1994ys}.
The Fisher matrix is given by
\begin{equation}
    \Gamma_{ij} = \left \langle \frac{\partial \tilde{h}(f)}{\partial \theta_i} \middle| \frac{\partial \tilde{h}(f)}{\partial \theta_j}  \right \rangle,
\end{equation}
where $\tilde{h}(f)$ is the Fourier transform of the time domain waveform $h(t)$, $\theta_i$, $i=1,2,...,$ are the parameters in $\tilde{h}(f)$. 
The inner product is defined as 
\begin{equation}
    \langle a | b \rangle  =4 \int^\infty_0 \frac{\tilde{a}(f) \tilde{b}^*(f) +\tilde{a}^*(f)\tilde{b}(f) }{2} \frac{df}{S_n(f)}
\end{equation}
where $*$ represents complex conjugate. 
The covariance matrix is related to the Fisher matrix through $\text{Cov}_{ij}=\Gamma^{-1}_{ij}$ \cite{maggiore2008gravitational}.
Therefore, the measurement error of a GW parameter is $\Delta \theta_i = \sqrt{\text{Cov}_{ii}}$.
For the sky localization error, it is given by 
\begin{equation}
    \Delta \Omega =2\pi |\sin \theta| \sqrt{ (\Delta \theta)^2 (\Delta \phi)^2 - (\Delta \theta \Delta \phi)^2 },
\end{equation}
where $\Delta \theta$, $\Delta \phi$, and $\Delta \theta \Delta \phi$ are given by the covariance matrix.

To quantify an effective GW signal, one needs the signal-to-noise ratio (SNR) $\rho$.
In line with the SNR threshold adopted by TianQin analysis \cite{Wang:2019ryf}, a GW detection is claimed only when the SNR is $\rho>8$. The SNR is given by
\begin{equation}
    \rho = \sqrt{ \langle \tilde{h} | \tilde{h} \rangle }.
\end{equation}
For multiple detectors, the combined SNR is simply the root sum square of component SNRs.

In this paper, we focus on the $\textit{TianQin I+II}$ configuration, which consists of twin satellite constellations with perpendicular orbital planes designed to avoid the 3-month gaps in data \cite{Liu:2020eko, Wang:2019ryf}.

\subsection{GW as standard siren}
Among different classes of standard sirens, bright sirens provide the most promising and direct way to obtain source redshift information \cite{Jin:2025dvf}, since the redshift can be determined from an identified electromagnetic (EM) counterpart or host galaxy.
Some previous studies have suggested that MBHB mergers may have observable EM signatures \cite{Armitage:2002uu, Jullo:2010dq, Gold:2014dta, Farris:2014zjo}, therefore MBHB mergers are potential bright siren candidates for GW cosmology studies \cite{Holz:2005df, Su:2024avk, Feng:2025wbz, Zhang:2026sxi, Song:2025bio}.
In this paper, we focus on the bright standard sirens from MBHB mergers.

For the MBH growth scenario, we mainly consider the light-seed scenario, with the heavy-seed scenario serving as a supplement.
In the light-seed scenario, we consider the population III (pop III) model, which assumes the MBH seed is the remnants of the first generation or population III stars, with a mass of around 100$M_\odot$ \cite{Madau:2001sc, Volonteri:2002vz}. 
In the heavy-seed scenario, we consider the Q3nod model, which assumes the MBH seeds are born from the direct collapse of protogalactic disks that my driven bar instabilities \cite{Klein:2015hvg}, with a mass of $\sim 10^5 M_\odot$.
For more details, we refer the reader to Ref.~\cite{Zhu:2021aat}.

Consider a MBHB of circular orbit at a luminosity distance $d_L$, with component masses $m_1$ and $m_2$, the frequency domain GW waveform can be expressed as \cite{Sathyaprakash:2009xs}
\begin{equation}
    \tilde{h} (f) = \frac{1}{D_{\text{eff}}} \sqrt{\frac{5}{96}}\frac{(G \mathcal{M}_c)^{5/6}}{\pi^{2/3} c^{3/2}} f^{-7/6} \text{exp}(- i \Phi (f; \mathcal{M}_c,\eta)),
\end{equation}
where $G$ is the gravitational constant, $\mathcal{M}=M \eta^{3/5}$ is the chirp mass, $\eta=m_1m_2/M^2$ is the symmetric mass ratio, $M=m_1+m_2$ is the total mass, and $\Phi (f; \mathcal{M}_c,\eta)$ is the phase of the waveform depending on the parameters $\mathcal{M}_c$ and $\eta$.
Note that the chirp mass $\mathcal{M}$ largely determines the overall evolution of the GW waveform, but the parameter directly measured from the data is the observed chirp mass $\mathcal{M}_{c} = (1+z) \mathcal{M}$.
In the following, we always simply refer chirp mass to the observed quantity $\mathcal{M}_{c}$.

The effective luminosity distance, $D_{\text{eff}}$, is defined as
\begin{equation}
    D_{\text{eff}} = d_L \left[ F_+^2 ( 1+\cos^2\iota )^2 + F^2_\times (2 \cos \iota)^2 \right]^{-1/2},
\end{equation}
where the antenna pattern functions $F_+$ and $F_\times$ can be found in Ref.~\cite{Feng:2019wgq}, and $d_L$ is the luminosity distance given by Eq.(\ref{eq:lum_dist}).
The background cosmological models mentioned in Section~\ref{sec:background-setup} enter the GW standard siren simulation through the luminosity distance $d_L$.

\subsection{GW errors and $\chi^2$ statistics}
Since GW detections allow us to directly measure the luminosity distances of GW sources, they can serve as the so-called "standard sirens" to independently constrain the cosmic expansion history. 
Therefore, the luminosity distance is the key quantity to construct the $\chi^2$ statistic function.
For a GW event, distance measurement usually has four kinds of uncertainties: the instrumental error $\sigma_{d_\text{L}}^{\text{inst}}$, weak lensing error $\sigma_{d_\text{L}}^{\text{lens}}$, peculiar velocity error $\sigma_{d_\text{L}}^{\text{pv}}$, and redshift measurement error $\sigma_{d_\text{L}}^{\text{reds}}$. 
For the instrumental error, suppose that the error of $d_L$ is uncorrelated with errors on the remaining GW parameters, we have
\begin{equation}
    \sigma_{d_\text{L}}^{\text{inst}}(z) \simeq \sqrt{\left \langle \frac{\partial \tilde{h}}{\partial d_L} \middle| \frac{\partial \tilde{h}}{\partial d_L}\right \rangle ^{-1}}.
\end{equation}
For the error caused by weak lensing, we use the fitting formula of Ref.~\cite{Hirata:2010ba} as:
\begin{equation}
    \sigma_{d_\text{L}}^{\text{lens}}(z)= d_\text{L} (z) \times 0.066 \left[ 4(1-(1+z)^{-0.25} \right]^{1.8}.
\end{equation}
For the error caused by the peculiar velocity of a GW source, a fitting formula given by Ref.~\cite{Kocsis:2005vv} is:
\begin{equation}
    \sigma_{d_\text{L}}^{\text{pv}}(z)= d_\text{L} (z) \times \left[ 1+ \frac{c(1+z)^2}{H(z) d_\text{L}(z)} \right] \frac{\sqrt{ \langle v^2 \rangle }}{c},
\end{equation}
where $\sqrt{ \langle v^2 \rangle }$ is the peculiar velocity of the GW source and is roughly fixed at 500 km/s. 
For the error caused by photometrical measurement, the redshift measurement error is estimated as \cite{Tamanini:2016zlh}:
\begin{equation}
    \sigma_{d_\text{L}}^{\text{reds}} = \frac{\partial d_L}{\partial z} (\Delta z)_n,
\end{equation}
where redshift uncertainty approximately as $(\Delta z)_n \simeq 0.03(1+z_n)$.
Therefore, the total error can be expressed as
\begin{equation}
    \sigma_{d_L}=\sqrt{(\sigma_{d_\text{L}}^{\text{inst}})^2+(\sigma_{d_\text{L}}^{\text{lens}})^2+(\sigma_{d_\text{L}}^{\text{pv}})^2 + (\sigma_{d_\text{L}}^{\text{reds}})^2 }.
\end{equation}
The $\chi^2$ function of GW data can be written as
\begin{equation}
    \chi^2_{\rm GW}=\sum^N_{i=1} \left[ \frac{d_L^{\text{i}} -d_L(z_i; \vec{\Omega} ) }{\sigma_{d_L}^i } \right]^2,
\end{equation}
where $z_i$, $d_L^i$, and $\sigma_{d_L}^i$ are the $i$th redshift, luminosity distance, and error of luminosity distance of GW data. $\vec{{\Omega}}$ represents the set of cosmological parameters.
Assuming the measurement errors be Gaussian, the likelihood function is given by
\begin{equation}
    \mathcal{L} \propto e^{-\chi^2_{\rm GW} /2}.
\end{equation}

\section{Results and Discussion}
\label{sec:results}
As mentioned above, by adopting the $\Lambda$CDM model and the wbin model as the background cosmological models, we obtained two mock TQSS catalogs, i.e., TQSS1 and TQSS2 catalogs.
In this Section, we first compare the differences in luminosity distance $d_L$ between TQSS1 and TQSS2 catalogs. 
Secondly, we investigate the cosmological implications of TQSS1 and TQSS2 catalogs.
Lastly, we discuss the impact of adopting different MBH seed models on the cosmological implications of TQSS mock catalogs.

\subsection{Differences in the luminosity distance}
\begin{figure*}[htbp]
    \centering
    \includegraphics[width=0.48\textwidth]{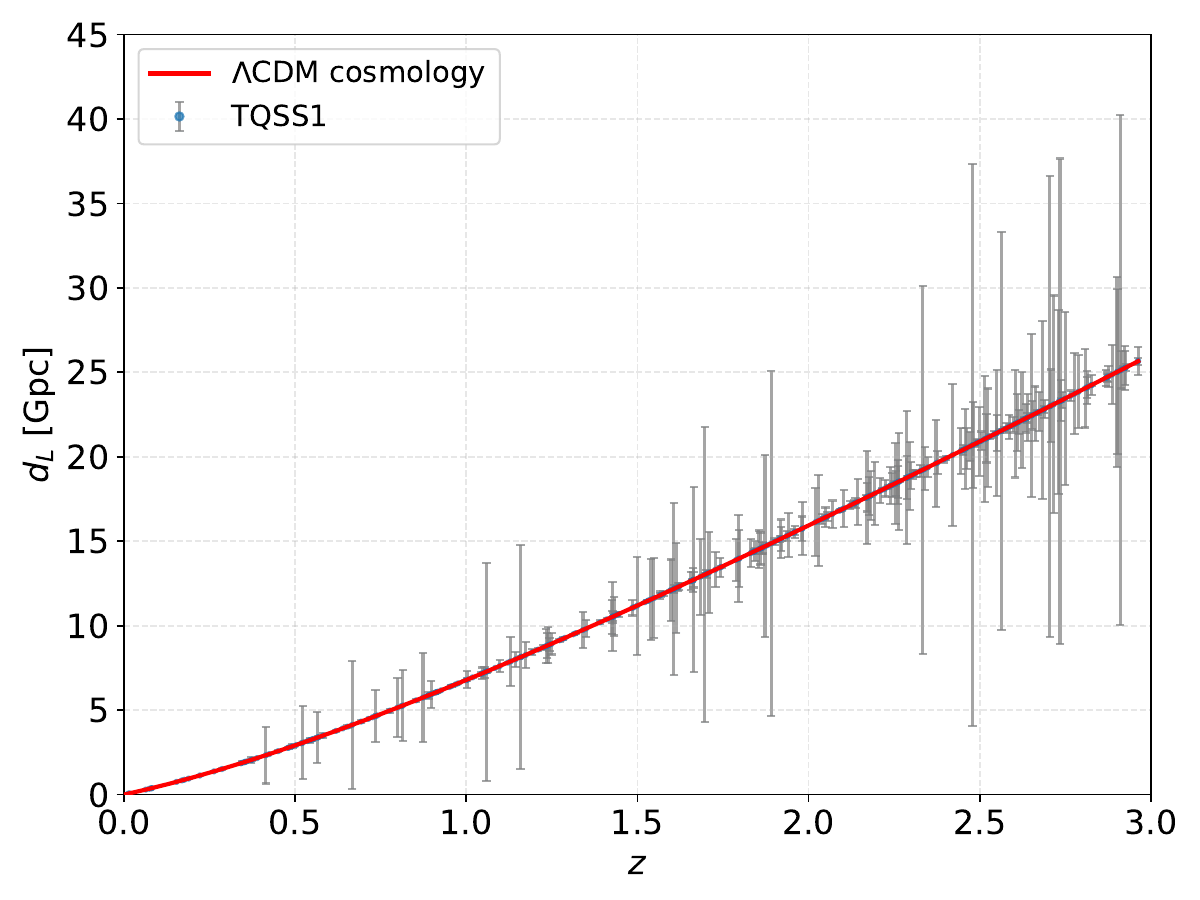}
    \includegraphics[width=0.48\textwidth]{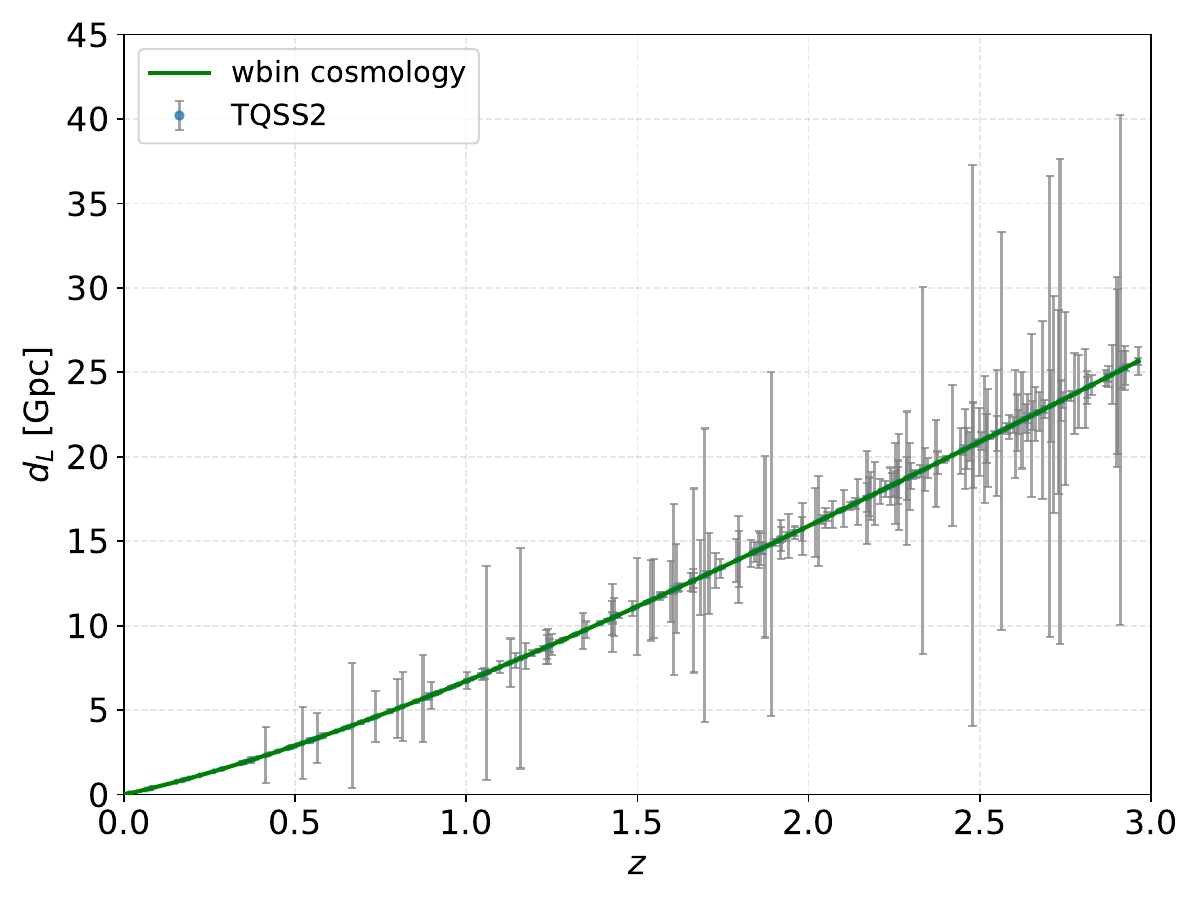}
    \caption{Mock TianQin standard siren catalog, where TQSS1 is generated based on the $\Lambda$CDM model and TQSS2 is generated based on the wbin model. The red and green solid lines represent the distance-redshift relation predicted by the $\Lambda$CDM cosmology and the wbin cosmology, respectively.}
    \label{fig:tqss}
\end{figure*}

\begin{figure}
\small
    \centering
    \includegraphics[width=0.48\textwidth]{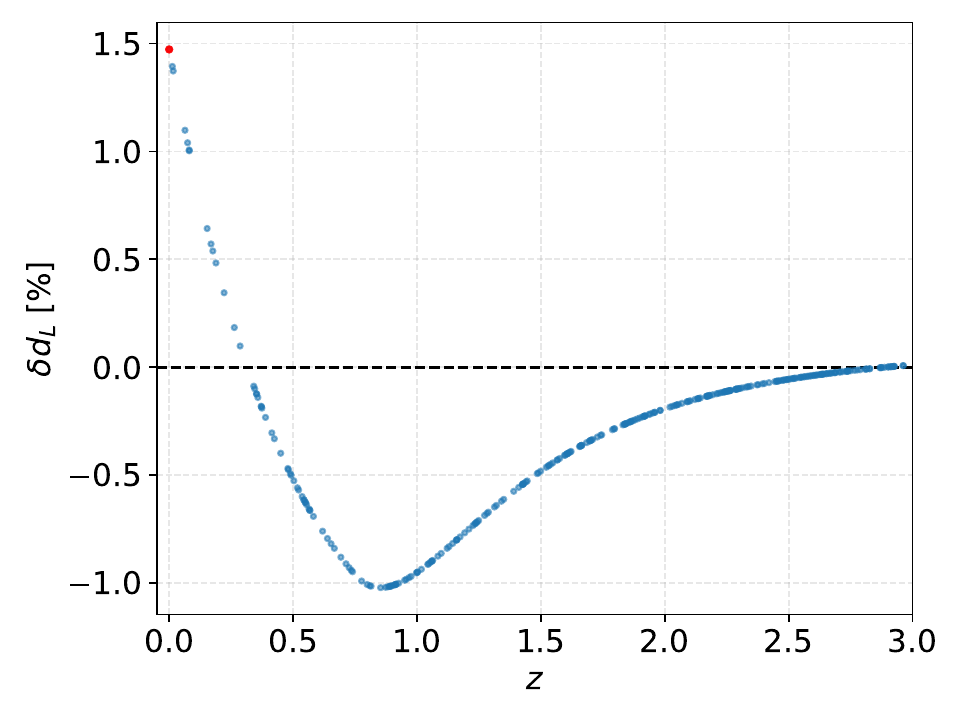}
    \caption{Fractional difference in the luminosity distance between TQSS1 and TQSS2 catalogs, calculated by $\delta d_L \equiv (d_L^{\rm TQSS2} - d_L^{\rm TQSS1})/d_L^{\rm TQSS1} $. The red dot in the upper-left corner marks the largest difference, reaching $1.48\%$.}
    \label{fig:DL_diff}
\end{figure}

In this subsection, we first present the simulated TQSS1 and TQSS2 catalogs in Fig.~\ref{fig:tqss}, and then we discuss the difference in the luminosity distance $d_L$, as shown in Fig.~\ref{fig:DL_diff}.

During the simulation, we consider that the GW signals are emitted by the MBHB mergers, which are treated as bright standard sirens.
The GW detections are based on the $\textit{TianQin I+II}$ configuration.
Adopting the pop III seed model, we follow Ref.~\cite{Zhu:2021aat} to generate 308 mock GW events from the MBHB mergers \footnote{The simulated TianQin standard siren catalogs are publicly available at \url{https://github.com/jxl2020/TQSS}. }.
The two mock catalogs differ only in their background cosmological models, while all other simulation settings are the same.

Figure~\ref{fig:tqss} shows the mock TQSS1 and TQSS2 catalogs.
The red and green curves represent the corresponding theoretical distance-redshift relations for the $\Lambda$CDM model and the wbin model, respectively.
For each simulated GW event, the blue dot denotes the mean value of luminosity distance, with the gray error bar indicating the $1\sigma$ uncertainty.

As shown in Fig.~\ref{fig:tqss}, one can hardly see the difference between TQSS1 and TQSS2 catalogs.
Therefore, we quantitatively investigate the difference of distance-redshift relation in Fig.~\ref{fig:DL_diff}.

Figure \ref{fig:DL_diff} presents the fractional difference of the luminosity distance between TQSS1 and TQSS2 catalogs.
The blue dots represent the fractional difference $\delta d_L$, which is calculated by $\delta d_L \equiv (d_L^{\rm TQSS2} - d_L^{\rm TQSS1})/d_L^{\rm TQSS1} $.
The largest difference is marked as a red dot in the upper-left corner of Fig.~\ref{fig:DL_diff}.
The fractional difference $\delta d_L$ first decreases over $0<z<0.32$, then increases over $0.32<z<0.85$, and finally decreases over $0.85<z<3.0$.

For $0<z<0.32$, $d_L^{\rm TQSS2}$ is greater than $d_L^{\rm TQSS1}$.
The largest difference occurs at the limit $z \to 0$, with a maximum difference of $1.48\%$.
We mark the largest difference as a red dot in the upper-left corner of Fig.~\ref{fig:DL_diff}.
It should be noted that the $1.48\%$ difference originates from the different input values of $H_0$ because $\lim\limits_{z\to 0} \delta d_L \simeq H_0^{\Lambda \rm CDM} / H_0^{\rm wbin} -1$.
Therefore, as the redshift approaches $0$, the contribution of the input $H_0$ values becomes increasingly dominant in the $ \delta d_L$.
The luminosity distances of two catalogs are equal at $z\simeq 0.32$, marking the transition from $d_L^{\rm TQSS2} > d_L^{\rm TQSS1}$ to $d_L^{\rm TQSS2} < d_L^{\rm TQSS1}$.
For $0.32<z<3$, $d_L^{\rm TQSS2}$ becomes smaller than $d_L^{\rm TQSS1}$.
On this occasion, the maximum difference reaches $1.02\%$ at $z= 0.85$.
This $1.02\%$ difference reflects the genuine difference in the DE EoS which underlies the background cosmological models.
As the redshift increases beyond $z=0.85$, the difference decreases monotonically and becomes negligible by $z\simeq 3$.
This is because DE is increasingly subdominant at high redshift.

In summary, we conclude that there are only small differences between TQSS1 and TQSS2 catalogs, with a maximum relative difference of $1.48\%$ in the distance-redshift relation.

\subsection{Cosmological implications}
In this subsection, we employ the CPL model to investigate the cosmological implications of the TQSS1 and TQSS2 catalogs.
We mainly focus on the cosmological constraint on the DE EoS parameters of the CPL model. 
In the first part, we present the cosmological results by using these two catalogs individually. 
In the second part, we combine these two catalogs individually with the combination of DESI DR2+CMB+DESY5.
The detailed results of the parameter constraints are summarized in Table \ref{tab:param}.
\begin{table*}
\small
    \centering
    \renewcommand{\arraystretch}{1.5}
    \setlength{\tabcolsep}{4pt}
    \begin{tabular}{lccccc}
    \hline \hline
    \textbf{Model/Dataset}   & $\Omega_m$  &  $H_0$[$\mathrm{km\,s^{-1}\,Mpc^{-1}}$]  & $w_0$ & $w_a$ \\
    \hline
      $\textbf{CPL}$ \\
      TQSS1     & $0.3006 ^{+0.0624}_{-0.0328}$ & $67.77_{-0.86}^{+0.87}$ & $-0.944_{-0.149}^{+0.126}$  & $-0.303^{+1.310}_{-0.613}$ \\
      TQSS2     & $0.2734^{+0.0843}_{-0.0429}$ & $66.63_{-0.75}^{+0.85}$ & $-0.766_{-0.124}^{+0.110}$  & $-0.281^{+1.082}_{-0.398}$ \\
      DESY5  & $0.3603 ^{+0.0849}_{-0.0263}$ & $74.85_{-9.16}^{+23.98}$ & $-0.807_{-0.109}^{+0.140}$  & $-1.572^{+0.465}_{-1.427}$ \\
      DESI DR2+CMB+DESY5           & $0.3199\pm0.0057$ & $66.65\pm0.56$ & $-0.766\pm0.058$  & $-0.790\pm0.226$\\
      DESI DR2+CMB+DESY5+TQSS1     & $0.3183\pm0.0043$ & $66.78\pm0.38$ & $-0.824\pm0.050$  & $-0.579\pm0.182$ \\
      DESI DR2+CMB+DESY5+TQSS2     & $0.3212\pm0.0044$ & $66.50\pm0.38$ & $-0.764\pm0.051$  & $-0.770\pm0.188$ \\
    \hline \hline
    \end{tabular}
    \caption{Summary table of cosmological parameter constraints. Results quoted for all parameters are the marginalized posterior means and $1\sigma$ confidence intervals in each case where two-sided constraints are possible.}
    \label{tab:param}
\end{table*}

\subsubsection{Constraints from individual catalogs}
In this part, we consider the case of using the TQSS1 and TQSS2 catalogs to constrain the CPL model.
For comparison, we also present the constraint result of DESY5 SN data alone.
\begin{figure*}
\small
    \centering
        \includegraphics[width=0.4\textwidth]{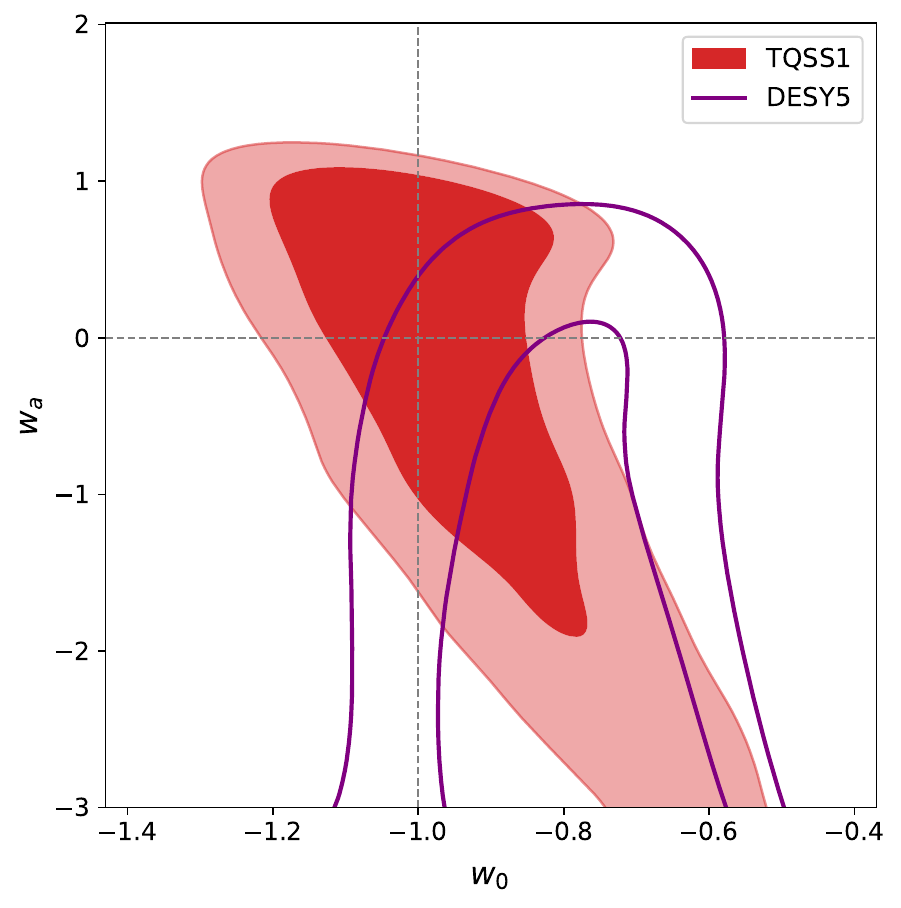}
        \includegraphics[width=0.4\textwidth]{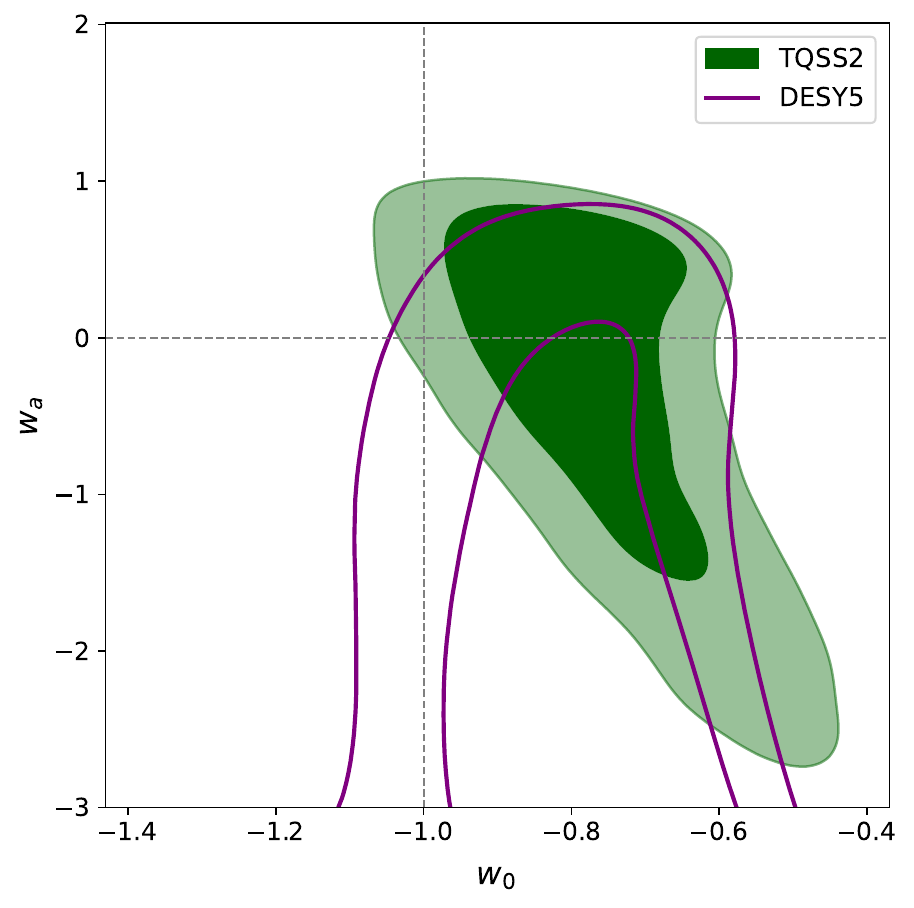}
    \caption{Two-dimensional marginalized contours in the $w_0$--$w_a$ plane at $1\sigma$ and $2\sigma$ confidence regions for the CPL model. Results from the TQSS1, TQSS2, and DESY5 data are in red, green, and purple, respectively.}
    \label{fig:contour_t1t2}
\end{figure*}

Figure \ref{fig:contour_t1t2} presents the two-dimensional marginalized contours in the $w_0$--$w_a$ plane at the $1\sigma$ and $2\sigma$ confidence regions for the CPL model.
In this figure, the red and green contours denote the constraint results from TQSS1 and TQSS2 catalogs alone, respectively. 
For comparison, the result from DESY5 alone is shown in purple.
As shown in Fig.~\ref{fig:contour_t1t2}, both the TQSS1 and TQSS2 catalogs place broad constraints on the $w_0$--$w_a$ plane, with areas similar to those inferred from DESY5.
When comparing the constraint results between the TQSS1 and TQSS2 catalogs, the TQSS2 catalog give a tighter constraint than the TQSS1 catalog.
Since the TQSS contour shapes and degeneracy directions are different from those of DESY5, the mock TQSS catalogs can provide an independent distance probe complementary to SN data.

\begin{figure}
\small
    \centering
        \includegraphics[width=0.48\textwidth]{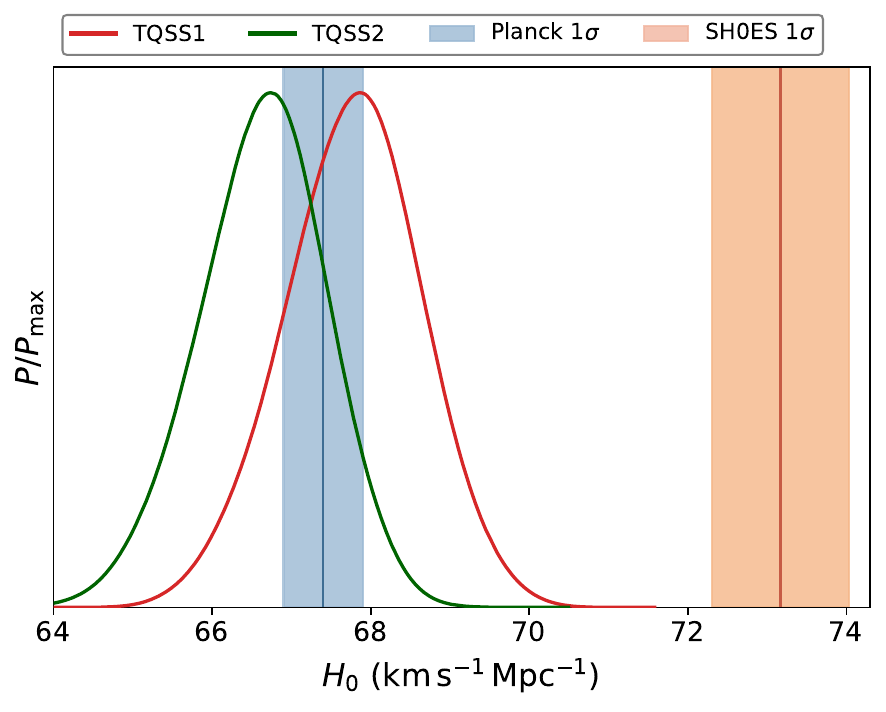}
    \caption{Normalized one-dimensional posterior distributions of the Hubble constant $H_0$ for the CPL model. The red and green curves denote the results of TQSS1 and TQSS2 catalogs, respectively. The light blue band denotes the $1\sigma$ confidence region of Planck, giving $H_0^{Planck}=67.4\pm 0.5$ $\rm km\,s^{-1}\,Mpc^{-1}$. The light orange band denotes the $1\sigma$ confidence region of SH0ES, giving $H_0^{\rm SH0ES}=73.17\pm 0.86$ $\rm km\,s^{-1}\,Mpc^{-1}$. }
    \label{fig:H0_t1t2}
\end{figure}

Figure \ref{fig:H0_t1t2} presents the normalized one-dimensional posterior distributions of the Hubble constant $H_0$ for the CPL model. 
The red and green curves denote the results of TQSS1 and TQSS2 catalogs, respectively. 
The light blue and light orange bands denote the $1\sigma$ confidence regions for the results of Planck and SH0ES, respectively. 
As shown in Fig.~\ref{fig:H0_t1t2}, the $H_0$ constraints from TQSS1 and TQSS2 catalogs favor lower values that are close to the value inferred from Planck data.
Taking the SH0ES measurement $H_0^{\rm SH0ES}=73.17\pm 0.86$ $\rm km\,s^{-1}\,Mpc^{-1}$ as a reference, the $H_0$ tension is $4.42\sigma$ for TQSS1 and is $5.56\sigma$ for TQSS2.
Since the simulation of the TQSS catalogs requires an input value of $H_0$, this process naturally affects the cosmological constraint on $H_0$ for the CPL model.
Therefore, the current mock TQSS catalogs cannot provide an independent assessment of the Hubble tension.

In summary, we conclude that either the TQSS1 or TQSS2 catalog alone gives cosmological constraints similar to those obtained from SN data, implying that these two catalogs can serve as a complement to mainstream cosmological observations.

\subsubsection{Constraints from data combinations}
In this part, we consider the case of combining the TQSS1 and TQSS2 catalogs with the DESI DR2+CMB+DESY5 combination.
For comparison, we also present the constraint result of DESI DR2+CMB+DESY5 alone.
Based on the joint constraints, we discuss the statistical significance of the preference for dynamical DE.
To provide a comprehensive analysis, we further investigate evolving behaviors of DE using three physical quantities: the DE EoS $w(z)$, the normalized DE density $f(z)$, and the deceleration parameter $q(z)$.

\begin{figure*}
\small
    \centering
        \includegraphics[width=0.4\textwidth]{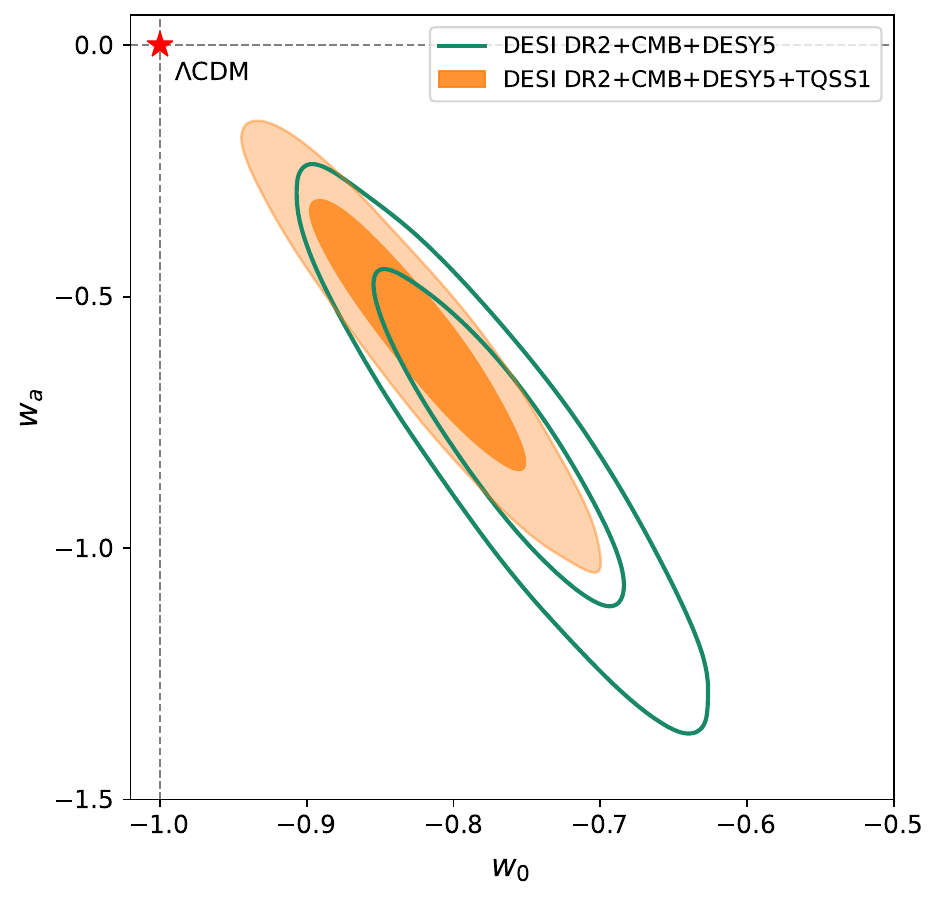}
        \includegraphics[width=0.4\textwidth]{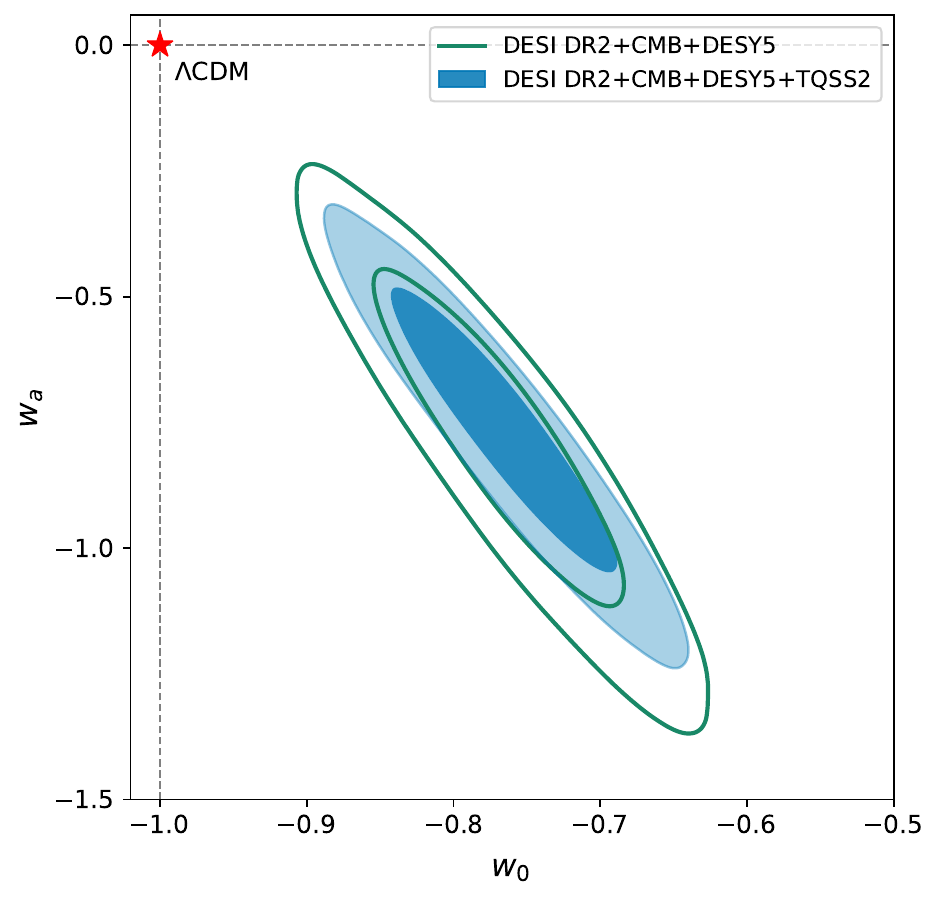}
    \caption{Two-dimensional marginalized contours in the $w_0$--$w_a$ plane at $1\sigma$ and $2\sigma$ confidence regions for the CPL model. The gray dashed lines indicate $w_0=-1$ and $w_a=0$; the red star denotes the $\Lambda$CDM limit ($w_0=-1$, $w_a=0$).}
    \label{fig:contour_w0wa}
\end{figure*}

Figure~\ref{fig:contour_w0wa} shows the two-dimensional marginalized contours in the $w_0$--$w_a$ plane for different data combinations. 
The orange and blue contours denote the constraint results from DESI DR2+CMB+DESY5+TQSS1 and DESI DR2+CMB+DESY5+TQSS2, respectively.
For comparison, the constraints from DESI DR2+CMB+DESY5 are shown as green contours.
The red star denotes the $\Lambda$CDM limit of the CPL model, corresponding to $w_0=-1$ and $w_a=0$.
As shown in Fig.~\ref{fig:contour_w0wa}, the $\Lambda$CDM limit lies outside the $2\sigma$ region for all data combinations, indicating a preference for dynamical DE.
When including the TQSS1 or TQSS2 catalogs in the DESI DR2+CMB+DESY5, the parameter space is further constrained.
Compared with the contour of DESI DR2+CMB+DESY5, the contour of DESI DR2+CMB+DESY5+TQSS1 is smaller and shifts slightly towards the $\Lambda$CDM limit.
As for the result of DESI DR2+CMB+DESY5+TQSS2, the corresponding contour is also smaller but shows no apparent shift.

We summarize the values of $\Delta \chi^2_{\rm MAP}$ and the corresponding statistical significances in Table \ref{tab:chi2}.
For the combination of DESI DR2+CMB+DESY5, the $\Delta \chi^2_{\rm MAP}$ value is $-11.6$, indicating a preference for an evolving DE EoS at $3.0\sigma$ CL.
When including TQSS1 in the DESI DR2+CMB+DESY5, the $\Delta \chi^2_{\rm MAP}$ increases to $-7.5$, lowering the preference to $2.3\sigma$ CL.
When including TQSS2 in the DESI DR2+CMB+DESY5, the $\Delta \chi^2_{\rm MAP}$ decreases to $-18.0$, strengthening the preference to $3.8\sigma$ CL.
The changes in $\Delta \chi^2_{\rm MAP}$ and significance are driven by the inclusion of different TQSS catalogs.
Since TQSS1 is generated based on the $\Lambda$CDM model, its inclusion in the DESI DR2+CMB+DESY5 combination naturally weakens the significance of the preference for dynamical DE.
In contrast, since TQSS2 is generated based on the wbin model, the corresponding result is more favorable to dynamical DE.
Therefore, the future TQSS could help clarify whether DE is evolving in the late-time Universe.

\begin{table}
\small
    \centering
    \renewcommand{\arraystretch}{1.5}
    \setlength{\tabcolsep}{4pt}
    \begin{tabular}{lcc}
    \hline \hline
    \textbf{Datasets}   & $\Delta \chi^2_{\rm MAP}$ &  Significance  \\
    \hline
      DESI DR2+CMB+DESY5             & $-11.6$  & $3.0\sigma$  \\
      DESI DR2+CMB+DESY5+TQSS1       & $-7.5$  & $2.3\sigma$    \\
      DESI DR2+CMB+DESY5+TQSS2       & $-18.0$ & $3.8\sigma$    \\
    \hline \hline
    \end{tabular}
    \caption{Summary of the difference in the effective $\chi^2_{\rm MAP}$ value, where $\Delta \chi^2_{\rm MAP}= \chi^2_{\rm CPL} - \chi^2_{\rm \Lambda CDM}$. }
    \label{tab:chi2}
\end{table}

\begin{figure*}[htbp]
\small
    \centering
        \includegraphics[width=\textwidth]{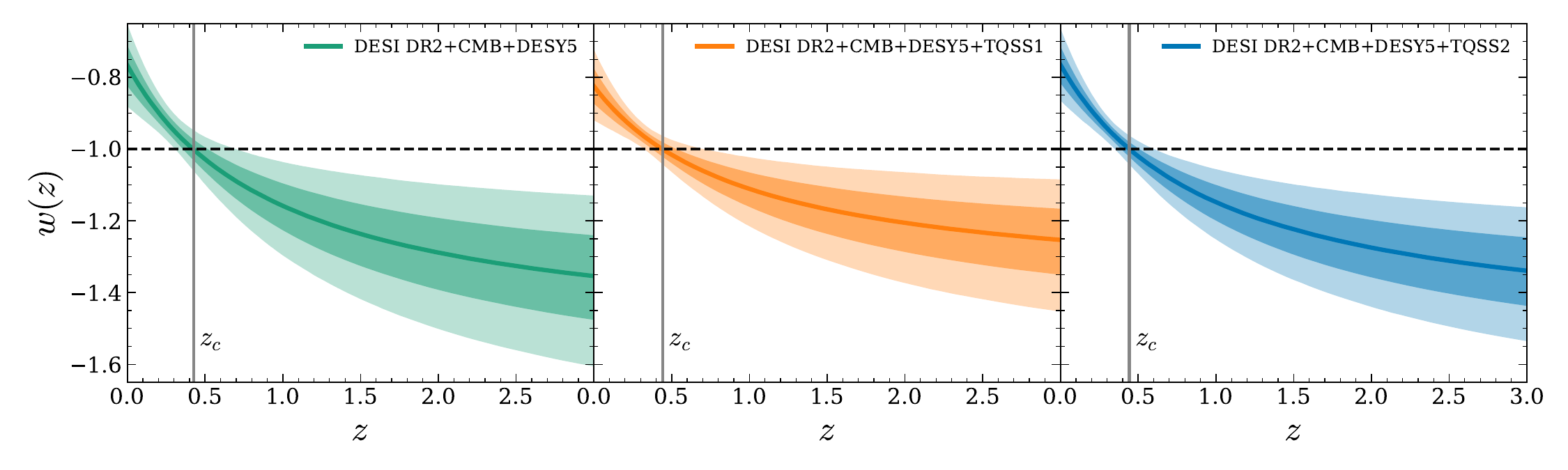}
        \includegraphics[width=\textwidth]{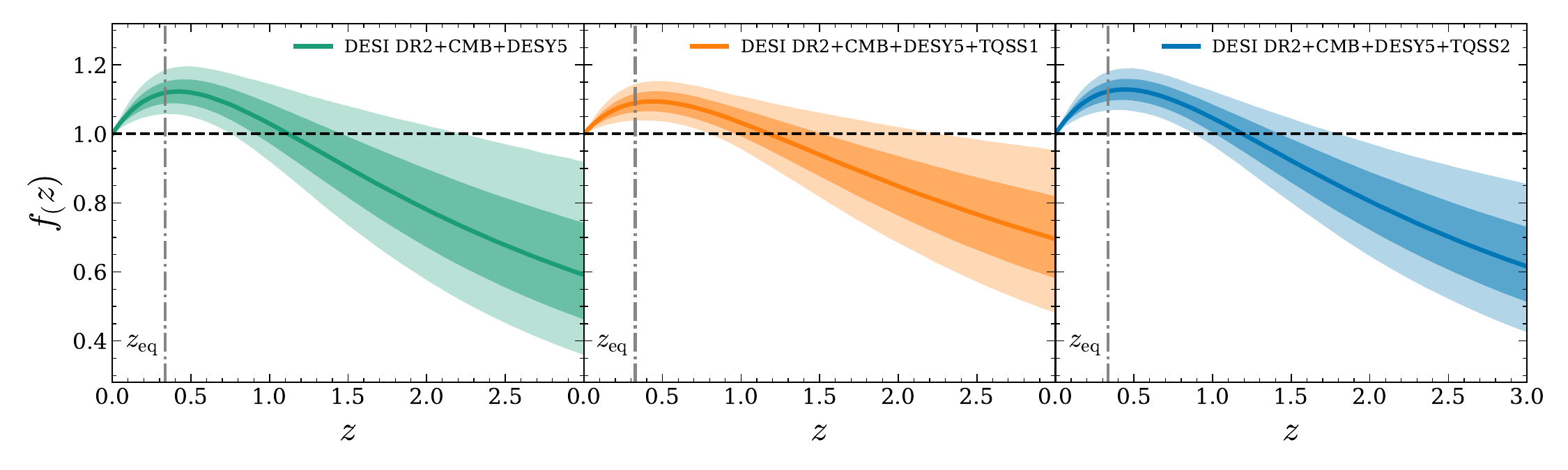}
        \includegraphics[width=\textwidth]{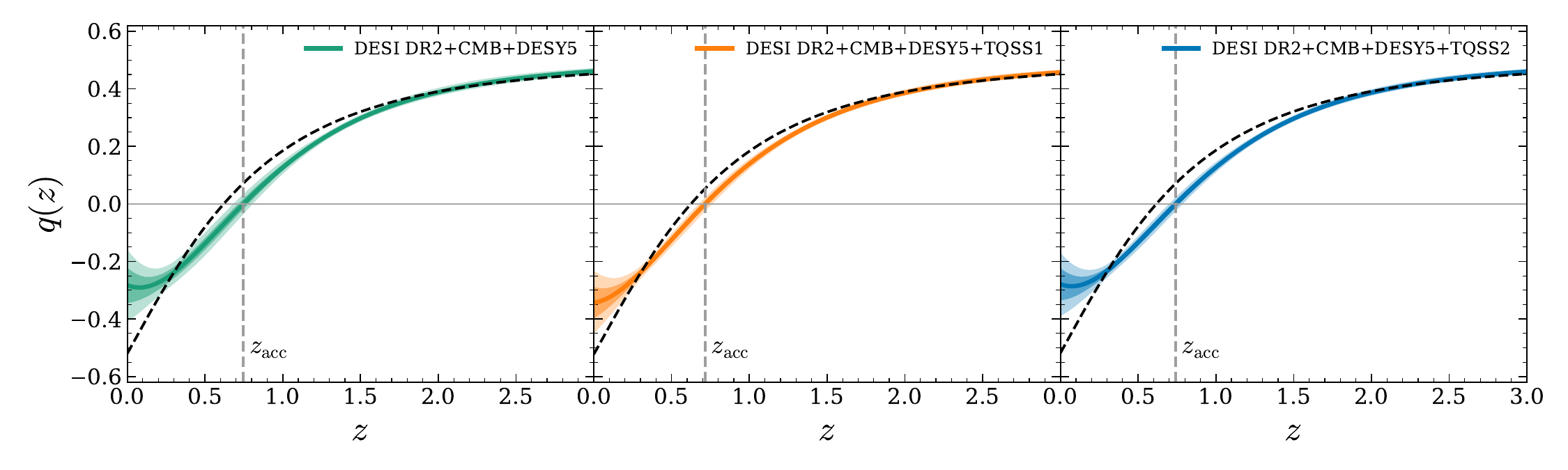}
    \caption{Evolution of the DE EoS $w(z)$, normalized DE density $f(z)$, and deceleration parameter $q(z)$ as a function of redshift in the CPL model. The colored lines and regions correspond to the mean, $1\sigma$, and $2\sigma$ confidence regions obtained from each data combinations. The black dashed lines represent predictions from the $\Lambda$CDM model. The solid, dashed-dotted, and dashed vertical gray lines indicate the phantom crossing redshifts $z_{\rm c}$, DE-matter equality redshifts $z_{\rm eq}$, and onset of cosmic acceleration redshifts $z_{\rm acc}$, respectively.}
    \label{fig:CPL_wfq}
\end{figure*}

In the following, we further explore the redshift evolution of DE using the three data combinations in Table \ref{tab:chi2}.
We consider three important physical quantities, including DE EoS $w(z)$, normalized DE density $f(z)$, and deceleration parameter $q(z)$. 
Specifically, $w(z)$, $f(z)$, and $q(z)$ are given by Eqs.(\ref{eq:cpl_w}), (\ref{eq:cpl_f}), (\ref{eq:cpl_q}), respectively.
We present these three quantities in Fig.~\ref{fig:CPL_wfq}.
The green lines denote the results from the data combination of DESI DR2+CMB+DESY5, while the orange and blue lines denote the results from DESI DR2+CMB+DESY5+TQSS1 and DESI DR2+CMB+DESY5+TQSS2, respectively.

In the top panel of Fig.~\ref{fig:CPL_wfq}, all the results of DE EoS $w(z)$ deviate from the $\Lambda$CDM case ($w=-1$).
This indicates that the EoS was phantom-like ($w(z)<-1$) in the past and has evolved to quintessence ($w(z)>-1$) at present.
The EoS crosses the phantom line ($w(z_{\rm c})=-1$) at $z_{\rm c}\simeq 0.42$, $0.43$, and $0.44$ for the three data combinations, respectively.

In the middle panel of Fig.~\ref{fig:CPL_wfq}, results of the normalized DE density $f(z)$ demonstrate that the density has a notable hump at low redshift and subsequently dilutes at high redshift.
The DE and matter densities are equal at $z_{\rm eq} \simeq 0.33$, $0.32$, and $0.33$ for the three data combinations, respectively.

In the bottom panel of Fig.~\ref{fig:CPL_wfq}, results of the deceleration parameter $q(z)$ suggest that the onset of cosmic acceleration ($q<0$) begins earlier than predicted by $\Lambda$CDM ($z_{\rm acc} \simeq 0.65$).
Moreover, the cosmic acceleration shows a slowing trend in recent times.
The cosmic acceleration begins at $z_{\rm acc}\simeq 0.75$, $0.71$, and $0.74$ for the three data combinations, respectively.

Overall, the inclusion of either TQSS1 or TQSS2 in the DESI DR2+CMB+DESY5 moderately tightens the confidence regions, further strengthening the evidence for evolving DE. 
The redshift evolution trend of $w(z)$, $f(z)$, and $q(z)$ remains unchanged.

In summary, we conclude that both TQSS1+DESI DR2+CMB+DESY5 and TQSS2+DESI DR2+CMB+DESY5 are still expected to favor dynamical DE.
Specifically, these two combinations showing a preference for dynamical DE at $2.3\sigma$ and $3.8\sigma$ CL, respectively.

\subsection{Impact of adopting different MBH seed models}
In the above analysis, the TQSS1 and TQSS2 catalogs are constructed based on the popIII light-seed model.
Recent studies suggest that the little red dots (LRDs) observed with the James Webb Space Telescope (JWST) may provide evidence for the formation of heavy black hole seeds \cite{Pacucci:2025ojp, Pacucci:2026ovn}.
Therefore, in this subsection, we further discuss the impact of adopting different MBH seed models on the cosmological implications of TQSS mock catalogs.

In the following, we consider the Q3nod heavy-seed model, and denote the corresponding catalogs as TQSS1(Q3nod) and TQSS2(Q3nod), respectively.
Since the Q3nod model predicts a higher MBHB detection rate than the popIII model, we follow Ref.~\cite{Zhu:2021aat} and generate 1045 mock GW events as bright standard sirens for each catalog.
After obtaining the new mock catalogs, we also employ the CPL model to investigate the cosmological implications of the TQSS1(Q3nod) and TQSS2(Q3nod) catalogs.

\begin{figure*}
\small
    \centering
    \includegraphics[width=0.4\textwidth]{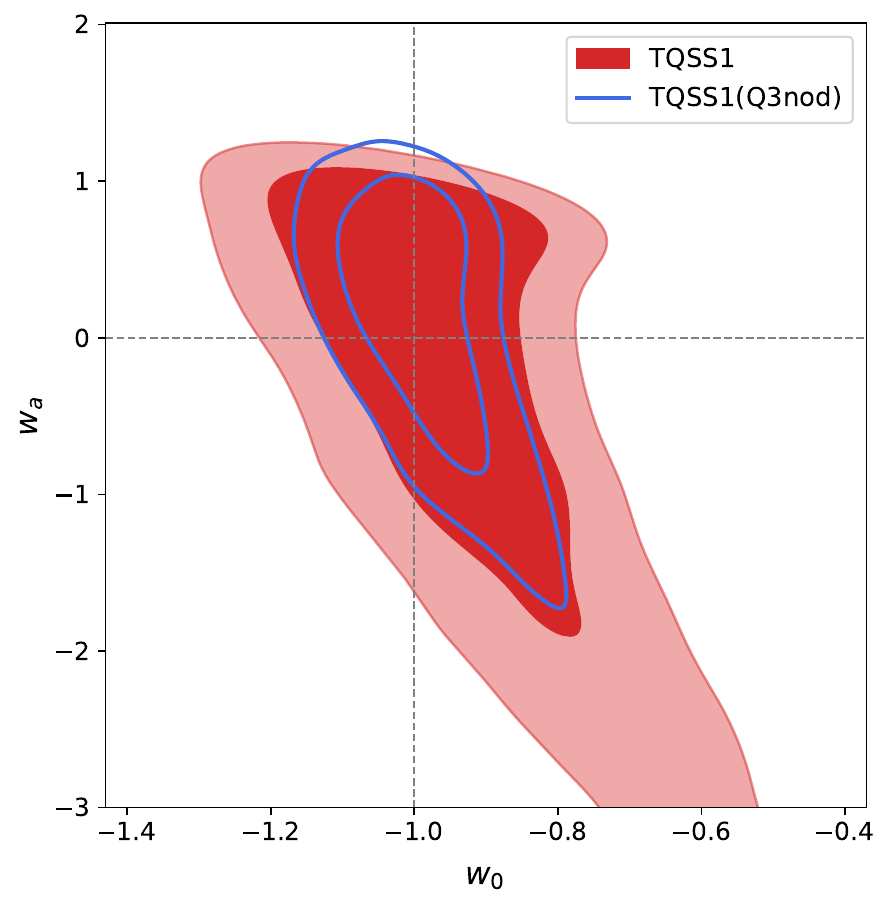}
    \includegraphics[width=0.4\textwidth]{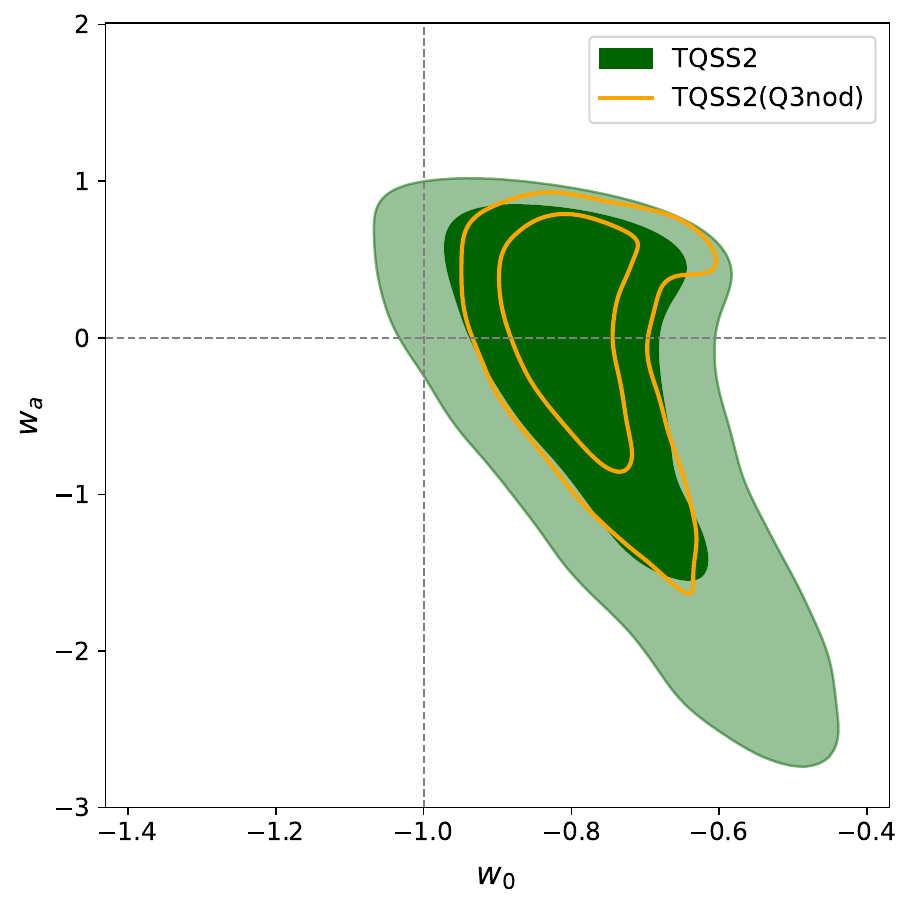}
    \caption{Two-dimensional marginalized contours in the $w_0$--$w_a$ plane at $1\sigma$ and $2\sigma$ confidence regions for the CPL model. Results from the TQSS1, TQSS1(Q3nod),  TQSS2, and TQSS2(Q3nod) are shown in red, blue, green and orange contours, respectively.}
    \label{fig:contour_TQSSq3d}
\end{figure*}

Figure~\ref{fig:contour_TQSSq3d} presents the two-dimensional marginalized contours in the $w_0$--$w_a$ plane at the $1\sigma$ and $2\sigma$ confidence regions for the CPL model.
In the left panel, the red and blue contours denote the constraint results for the TQSS1 and TQSS1(Q3nod) catalogs, respectively.
In the right panel, the green and orange contours denote the constraint results for the TQSS2 and TQSS2(Q3nod) catalogs, respectively.
As shown in Fig.~\ref{fig:contour_TQSSq3d}, compared with the popIII-based catalogs, the TQSS1(Q3nod) and TQSS2(Q3nod) catalogs give tighter constraints in the $w_0$--$w_a$ plane. 
This is because the Q3nod-based catalogs contain more GW events.
For constraints on the cosmological parameters, the TQSS1(Q3nod) catalog gives $\Omega_m=0.2883^{+0.0475}_{-0.0282}$, $H_0=68.01_{-0.44}^{+0.53}$ $\mathrm{km\,s^{-1}\,Mpc^{-1}}$, $w_0=-0.989_{-0.070}^{+0.067}$, and $w_a=0.128^{+0.822}_{-0.342}$, while the TQSS2(Q3nod) catalog gives $\Omega_m=0.2537^{+0.0762}_{-0.0354}$, $H_0=66.80_{-0.40}^{+0.51}$ $\mathrm{km\,s^{-1}\,Mpc^{-1}}$, $w_0=-0.788_{-0.072}^{+0.057}$, and $w_a=0.027^{+0.704}_{-0.272}$.
Overall, adopting the Q3nod heavy-seed model mainly increases the number of GW events and thus provides tighter constraints.
The contours area remain consistent with those obtained in the popIII case, indicating that our conclusions are robust against the choice of MBH seed models.

\begin{figure*}
\small
    \centering
    \includegraphics[width=0.4\textwidth]{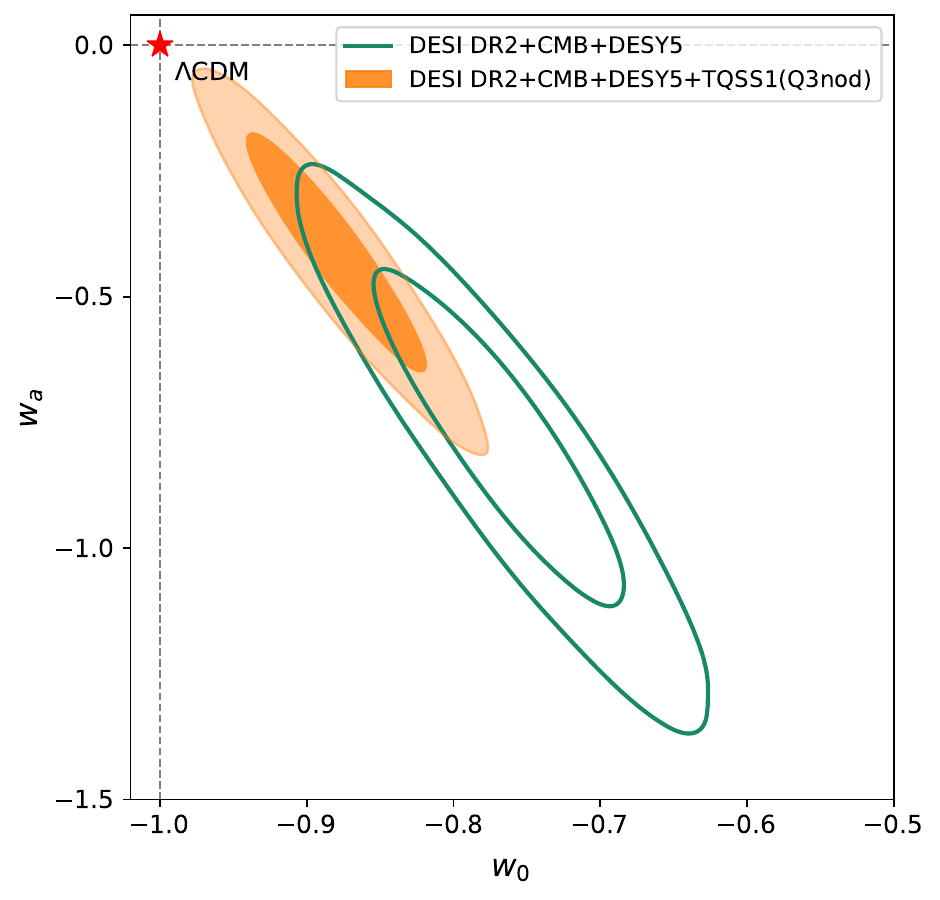}
    \includegraphics[width=0.4\textwidth]{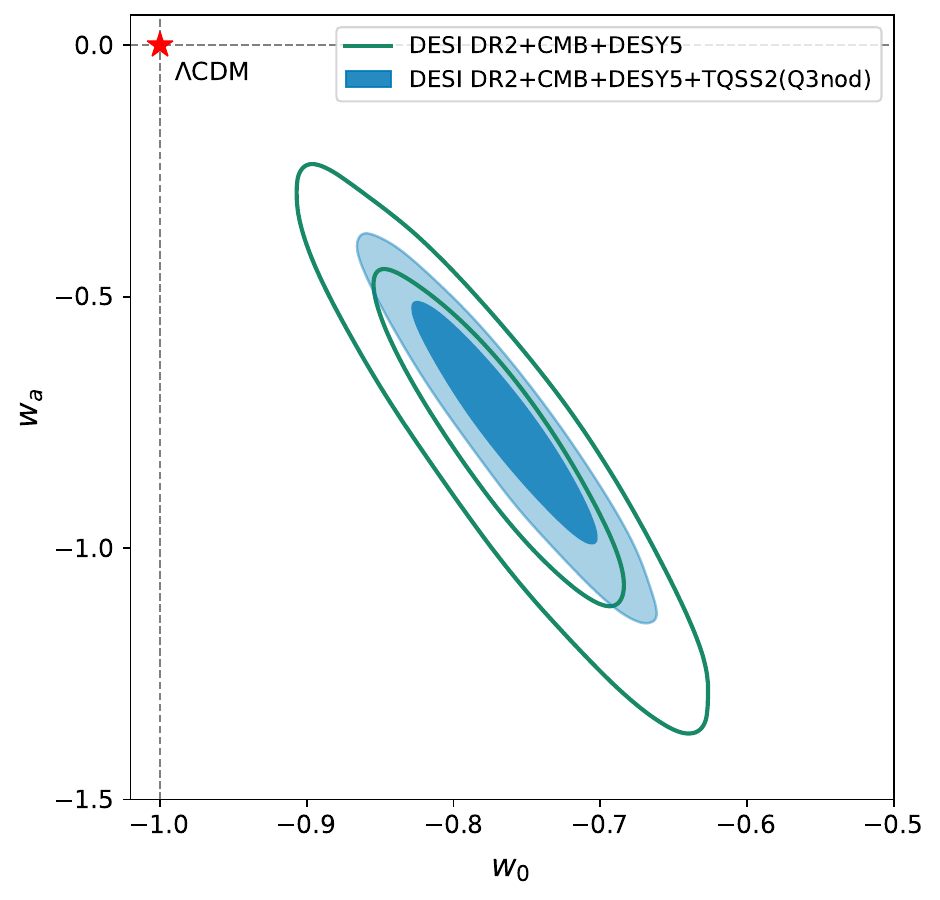}
    \caption{Two-dimensional marginalized contours in the $w_0$--$w_a$ plane at $1\sigma$ and $2\sigma$ confidence regions for the CPL model. The gray dashed lines indicate $w_0=-1$ and $w_a=0$; the red star denotes the $\Lambda$CDM limit ($w_0=-1$, $w_a=0$). }
    \label{fig:contour_bcst_q3d}
\end{figure*}

Figure~\ref{fig:contour_bcst_q3d} shows the two-dimensional marginalized contours in the $w_0$--$w_a$ plane for different data combinations. 
The orange and blue contours denote the constraint results for the DESI DR2+CMB+DESY5+TQSS1(Q3nod) and DESI DR2+CMB+DESY5+TQSS2(Q3nod) data combinations, respectively.
For comparison, the constraints for the DESI DR2+CMB+DESY5 data combination are shown as green contours.
The red star denotes the $\Lambda$CDM limit of the CPL model, corresponding to $w_0=-1$ and $w_a=0$.
As shown in Fig.~\ref{fig:contour_bcst_q3d}, the $\Lambda$CDM limit lies outside the $2\sigma$ region for all data combinations, indicating a preference for dynamical DE.
This is similar to the results in Fig.~\ref{fig:contour_w0wa}.
Adding the TQSS1(Q3nod) or TQSS2(Q3nod) catalog to DESI DR2+CMB+DESY5 further tightens the constraints.
Compared with the DESI DR2+CMB+DESY5 contour, the DESI DR2+CMB+DESY5+TQSS1(Q3nod) contour is smaller and shifts noticeably towards the $\Lambda$CDM limit.
In contrast, the DESI DR2+CMB+DESY5+TQSS2(Q3nod) contour is also smaller but shows no apparent shift.
As for the corresponding cosmological parameters, the DESI DR2+CMB+DESY5+TQSS1(Q3nod) combination gives $\Omega_m=0.3145\pm0.0036$, $H_0=67.15\pm0.28$ $\mathrm{km\,s^{-1}\,Mpc^{-1}}$, $w_0=-0.879\pm0.040$, and $w_a=-0.414\pm0.156$, while the DESI DR2+CMB+DESY5+TQSS2(Q3nod) combination gives $\Omega_m=0.3213\pm0.0036$, $H_0=66.44\pm0.28$ $\mathrm{km\,s^{-1}\,Mpc^{-1}}$, $w_0=-0.763\pm0.041$, and $w_a=-0.754\pm0.159$.
These two combinations show similar constraint precision but differ noticeably in the DE EoS parameters.
Since the TQSS catalogs drive the constraints in opposite directions depending on the underlying cosmology, future TianQin standard siren observations could help clarify whether DE is evolving in the late-time Universe.

In summary, adopting different MBH seed models does not impact the cosmological implications of the mock TQSS catalogs.

\section{Summary}
\label{sec:Summary}
Future space-borne GW detectors are anticipated to detect GW standard sirens for cosmological parameter estimation.
In this paper, we mainly focus on the cosmological implications of standard sirens from the TianQin GW detector.
All previous forecasts of TQSS adopted the $\Lambda$CDM model to generate mock catalogs.
However, the $\Lambda$CDM model faces two serious observational challenges:
the Hubble tension problem and the preference for dynamical DE indicated by DESI BAO measurements.
Therefore, we adopt a redshift-binned DE reconstruction model (the wbin model) as the background cosmological model to generate the mock TQSS catalog.
This choice has two main advantages: 
(1) it does not assume a specific functional form for the DE EoS; 
(2) it can reveal the details of DE's dynamical evolution.
The use of the wbin model to generate TQSS catalogs is the biggest novelty of this paper.

In order to compare the impact of the background cosmology, we consider two models: the $\Lambda$CDM model and the wbin model.
Since the simulation of GW data requires cosmological parameters as input, we constrain these two models using DESI DR2 BAO data, $\textit{Planck}$ 2018 CMB data, and DESY5 SN data.
Note that we employ the full CMB likelihood rather than the CMB distance priors for the CMB data.
We adopt the popIII light-seed model for MBHs and consider the bright standard sirens from MBHB mergers.
Making use of the two background cosmological models and the obtained cosmological parameters, we generate two mock TQSS catalogs. 
The TQSS1 catalog is based on the $\Lambda$CDM model and the TQSS2 catalog is based on the wbin model.
We first compare the distance-redshift relations of these two catalogs, and then investigate their cosmological implications for dynamical DE under the CPL model.
We analyze the TQSS1 and TQSS2 catalogs both individually and in combination with DESI DR2+CMB+DESY5.
In addition, to further discuss the impact of adopting different MBH seed models on the TQSS catalogs, we consider the Q3nod heavy-seed model for comparison.

We find that there are only small differences between TQSS1 and TQSS2 catalogs, with a maximum relative difference of $1.48\%$ in the distance-redshift relation.
In addition, it is found that either the TQSS1 or TQSS2 catalog alone gives cosmological constraints similar to those obtained from SN data, implying that these two catalogs can serve as a complement to mainstream cosmological observations. 
Most importantly, we find that both TQSS1+DESI DR2+CMB+DESY5 and TQSS2+DESI DR2+CMB+DESY5 are still expected to favor dynamical DE, with the two combinations showing a preference for dynamical DE at $2.3\sigma$ and $3.8\sigma$ CL, respectively.
It should be noted that adopting different massive black hole seed models does not impact the cosmological implications of the mock TQSS catalogs.

In this paper, we quantitatively assessed the differences between TQSS1 and TQSS2 catalogs and investigated their cosmological implications.
However, we only considered bright standard sirens from MBHB mergers in simulating the TQSS catalogs.
Therefore, future studies could consider dark standard sirens from SBHB and EMRI systems.
In addition, since we only adopted the CPL model to discuss the dynamical evolution of DE, it would be worthwhile to use the mock TQSS catalogs to constrain other DE models, including early DE models, interacting DE models, etc.
Moreover, another promising direction is to investigate the potential of TQSS catalogs to shed light on the Hubble tension problem and the cosmic age problem.
These topics are worth further exploration in future work.

\begin{acknowledgments}
We thank Tian-Cheng Luan at Sun Yat-Sen University for providing technical support in using the High Performance Computing resources.
S.Wang is supported by a grant of Guangdong Provincial Department of Science and Technology under No. 2020A1414040009.
\end{acknowledgments}

\bibliography{apssamp}
\bibliographystyle{utphys}

\end{document}